\documentclass[12pt,eqsecnum,nofootinbib,floats,preprint,aps,prd,floatfix,titlepage,superscriptaddress,tightenlines]{revtex4} 
\usepackage{epsfig}
\usepackage{amsmath}
\usepackage{amssymb}
\usepackage{graphics}
\usepackage{bm}
\usepackage{color}

\newcommand{\bel}[1] {\begin{equation}\label{#1}}
\newcommand{\beal}[1] {\begin{eqnarray}\label{#1}}
\def\be{\begin{equation}}
\def\ee{\end{equation}}

\begin{document}
\title{Tunneling from a Minkowski vacuum to an AdS vacuum: A new thin-wall regime}
\author{Ali Masoumi}
\email{ali@cosmos.phy.tufts.edu}
\affiliation{Institute of Cosmology, Department of Physics and Astronomy,
Tufts University, Medford, MA 02155, USA}
\author{Sonia Paban}
\email{paban@zippy.ph.utexas.edu}
\affiliation{Department of Physics and Texas Cosmology Center
The University of Texas at Austin, TX 78712, USA}
\author{Erick J. Weinberg}
\email{ejw@phys.columbia.edu}
\affiliation{Physics Department, Columbia University, New York, New York 10027, USA}

\begin{abstract}

Using numerical and analytic methods, we study quantum tunneling from
a Minkowski false vacuum to an anti-de Sitter true vacuum.  Scanning
the parameter space of theories with quartic and non-polynomial
potentials, we find that for any given potential tunneling is
completely quenched if gravitational effects are made sufficiently
strong.  For potentials where $\epsilon$, the energy density
difference between the vacua, is small compared to the barrier height,
this occurs in the thin-wall regime studied by Coleman and De Luccia.
However, we find that other potentials, possibly with $\epsilon$ much
greater than the barrier height, produce a new type of thin-wall bounce
when gravitational effects become strong.  We show that the critical
curve that bounds the region in parameter space where the false vacuum
is stable can be found by a computationally simple
overshoot/undershoot argument.  We discuss the treatment of boundary
terms in the bounce calculation and show that, with proper regularization,
one obtains an identical finite result for the tunneling exponent
regardless of whether or not these are included.  Finally, we briefly
discuss the extension of our results to transitions between anti-de
Sitter vacua.

\end{abstract}

\preprint{UTTG-02-16}
\maketitle

\section{Introduction}

The possibility of transitions between field theory vacua in curved
spacetime has been a subject of considerable interest ever since the
seminal paper of Coleman and De Luccia (CDL)~\cite{Coleman:1980aw}
incorporated gravitational effects into the bounce
formalism~\cite{Coleman:1977py,Callan:1977pt,Weinberg:2012pjx}.  More
recently the possibility of a string theory landscape has brought
renewed interest in the subject.

Considerable attention has been focused on transitions between de
Sitter vacua.  These could well have occurred in our past.  They could
also be part of an eternal future, thanks in part to the fact that the
thermal nature of de Sitter space allows one to tunnel
upward~\cite{Lee:1987qc} as well as downward.  Somewhat less attention
has been paid to transitions that end in an anti-de Sitter (AdS)
vacuum.  Because one cannot tunnel upward from an AdS vacuum, such
transitions cannot be part of our past, but could be part of our
future.  If they are, it will be a grim future, because a bubble of
AdS space soon develops a
singularity~\cite{Coleman:1980aw,Abbott:1985kr}.  

However, even if there are AdS vacua, we may be protected against this
fate.  Working in the context of the thin-wall approximation, CDL
found that gravitational effects reduce the rate for tunneling from a
Minkowski vacuum to an AdS vacuum; if the difference in vacuum energy
densities is sufficiently small, the decay is completely
forbidden~\cite{Coleman:1980aw,Abbott:1985qa,Park:1986xa}.  This result can be
  extended to transitions between AdS vacua~\cite{Parke:1982pm}.  If
  the initial vacuum is de Sitter, tunneling is suppressed, although
  complete suppression is only possible in the flat-space limit of
  vanishing vacuum energy density~\cite{Bousso:2006am,Aguirre:2006ap}.

Decays to AdS vacua beyond the thin-wall limit were studied by Samuel
and Hiscock~\cite{Samuel:1991dy} and
others~\cite{Banks:2002nm,Banks:2009nz,Kanno:2011vm,Kanno:2012zf,Lee:2014ufa,
Espinosa:2015zoa,Branchina:2016bws}.
In this paper we go beyond these works by considering a wider range of
potentials and parameters.  By doing so we uncover some apparently
universal properties of these transitions that were not previously
recognized.  Perhaps most notably, we find a strong gravity regime in
which a new type of thin-wall bounce emerges.

We focus on transitions from a Minkowski false vacuum with vanishing
cosmological constant to an AdS vacuum with negative energy density
$-\epsilon$.  The case where $\epsilon$ is small compared to the
height $U_{\rm top}$ of the potential barrier between the false vacuum
and the AdS true vacuum is amenable to the thin-wall approximation,
and it was this case that was studied in detail by CDL.  We will be
concerned with the more general case, including theories where, at
least in the absence of gravity, the transition is mediated by
Euclidean solutions that are very definitely thick-wall bounces.

Although we restrict ourselves to theories with a single scalar field,
we expect our results to generalize to more complex theories,
including those where suppression of vacuum decay is due to 
supersymmetry, either exact or nearly  
so~\cite{Weinberg:1982id,Ceresole:2006iq,Dine:2007er,Dine:2009tv}.

In the usual, small $\epsilon$, thin-wall approximation the bounce
consists of a thin spherical shell that separates an interior true
vacuum region from a false vacuum exterior.  If the potential is
modified so that $\epsilon$ is decreased, the thickness of the shell
is unchanged, while the bounce radius grows.  The thin wall bounces
that we find arise when the gravitational constant $\kappa= 8 \pi G$ is
increased while the potential $U(\phi)$ is held fixed (or,
equivalently, if all mass scales in the potential are increased by
a uniform factor with $\kappa$ kept constant).  They consist of a spherical
region of pure AdS true vacuum that is enclosed by a transition region
that is composed of an inner part with a spatially varying negative
energy density and an outer part where the matter field traverses the
potential barrier.  When $\kappa$ is increased, the field profile in
the transition region and the thickness of this region (measured in an
appropriate coordinate) vary only weakly, while the radius of the true
vacuum region grows and becomes infinite at a potential-dependent
critical value $\kappa_{\rm cr}$.  The bounce action diverges as this
critical value is approached, so that tunneling is completely quenched
for $\kappa \ge \kappa_{\rm cr}$.

The remainder of the article is arranged as follows.  In
Sec.~\ref{setup-section} we review the bounce formalism for vacuum decay,
including the usual thin-wall approximation.  We also review some
energy considerations that play a role in the suppression of vacuum
decay at large $\kappa$.  In Sec.~\ref{thickwall-sec} we present
numerical results for bounces in both quartic polynomial and
non-polynomial potentials, with a range of both the parameters in the
potentials and of values of $\kappa$.  In Sec.~\ref{analytic-sec} we
analyze these numerical results and present analytic arguments that
explain some of the properties that we have uncovered.
Section~\ref{conclusion} contains some concluding remarks.  There are
three appendices.  Appendix~\ref{boundary-app} discusses possible
boundary terms in bounce calculations, and shows that their inclusion
or omission, if done properly, has no effect on the decay rate.
Appendix~\ref{numerical-app} gives details of our numerical methods.
Appendix~\ref{Ads-app} discusses the extension of our results to 
decays from one AdS vacuum to another.

\section{Set up and Summary of Previous Results}
\label{setup-section}

\subsection{Basic formalism}

\begin{figure}
   \centering
   \includegraphics[width=3.2in]{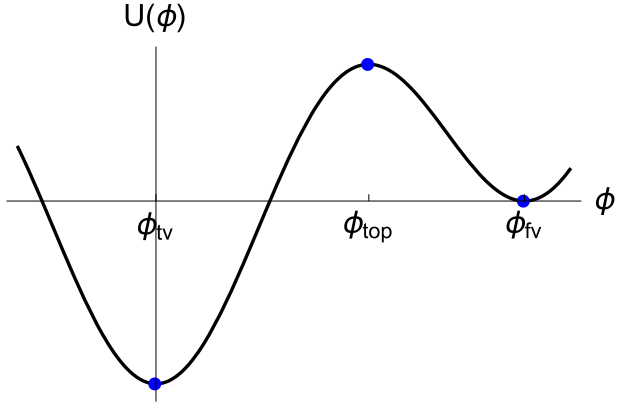}
   \caption{A quartic potential with two minima. }
   \label{potential-points}
\end{figure}

In this paper we will be concerned with vacuum transitions in a theory
with the dynamics of a scalar field $\phi$ governed by a potential
such as that shown in Fig.~\ref{potential-points}.  Without loss of
generality we have set the true vacuum at $\phi_{\rm tv}=0$ and put
the false vacuum to its right, at $\phi_{\rm fv}\equiv v >0$.
The peak of the potential barrier separating the
two vacua is at $\phi_{\rm top}$.  The values of the potential at these
points are $U_{\rm tv} = -\epsilon$, $U_{\rm fv} = 0$, and $U_{\rm top}$,
respectively.  It is important to note that $\epsilon$ is not necessarily 
small; indeed, we will consider some cases in which it is rather large.

An initial false vacuum state can decay by the nucleation of bubbles 
of true vacuum, with a nucleation rate per unit volume that can be 
written in the form
\begin{equation}
   \frac{\Gamma}{\cal{V}} = A e^{-B} \, .
\label{Gamma-form}
\end{equation}
In the absence of gravity, this nucleation rate can be calculated with
the aid of a bounce solution $\phi_b(x)$ of the Euclidean field
equations that begins with a configuration of spatially homogeneous
false vacuum at Euclidean time $x_4 = -\infty$, evolves to a
configuration containing a true vacuum bubble at $x_4=0$, and then
returns in an $x_4$-reversed fashion to a false vacuum configuration
at $x_4 =
\infty$~\cite{Coleman:1977py,Callan:1977pt}.
The tunneling exponent
\begin{equation}
   B = S(\phi_b)-S(\phi_{\rm{fv}})
\end{equation}
is the difference between the Euclidean action of the bounce and that
of the pure false vacuum\footnote{Because all actions in this paper
  will be Euclidean, we will simplify notation by omitting a subscript
  $E$.}.  The Lorentzian configuration at nucleation is given by a
spacelike slice through the bounce at $x_4=0$; at nucleation the
Lorentzian-time field is instantaneously at rest.

Coleman and De Luccia argued that the effects of gravity 
on bubble nucleation could be incorporated 
by adding a gravitational contribution to the 
action, which for a theory with a single scalar field then 
reads
\begin{equation}
   S=\int d^4x ~\sqrt{g}\left[ 
   \frac{1}{2} g^{ab} \partial_a \phi\, \partial_b \phi + U(\phi) 
  -\frac{1}{2 \kappa} R \right] + S_{\rm{bdy}} \,.
\label{fullaction}
\end{equation}
Here the last term in the integral is the Einstein-Hilbert action, with 
$\kappa = 8\pi G_N$, while
$S_{\rm{bdy}} $ is the Euclidean version of the Gibbons-Hawking
boundary action \cite{Gibbons:1976ue}.   

This boundary term, which does not appear explicitly in Ref.~\cite{Coleman:1980aw},
requires some comment.  Decays from a de Sitter vacuum are governed by
compact bounces, so there is no boundary term. On the other hand,
the bounces for decays from Minkowski or anti-de Sitter vacua are
infinite in extent, so their actions are potentially divergent, as are
the Gibbons-Hawking terms, and must be regularized.  The physically
significant quantity is the difference $B$ between the actions of the
bounce and of the pure false vacuum.  We show in
Appendix~\ref{boundary-app} that with proper regularization $B$ is
finite and independent of whether or not the Gibbons-Hawking term is
included.

For a single scalar field, it is known that in flat spacetime the
bounce has O(4) symmetry~\cite{Coleman:1977th}. Although no comparable result has been
proven for curved spacetime, in this work we will assume the same
symmetry. With this symmetry the metric takes the form:
\begin{equation} 
  ds^2= d\xi^2 + \rho(\xi)^2 d\Omega_3^2 \,.
\end{equation} 
The action can then be written as 
\begin{equation}
 S=2 \pi^2 \,\int_{\xi_{min}}^{\xi_{max}} d\xi \left\{ \rho^3
  \left[ \frac{1}{2} \phi'^2 + U(\phi) \right]
  + \frac{3}{\kappa} (\rho^2 \rho''+ \rho \rho'^2 - \rho)\right\}
    - \left.\frac{6 \pi^2}{\kappa}\,
   \rho^2 \rho'\right|^{\xi=\xi_{\rm max}}_{\xi={\xi_{min}}}
\label{action1}
\end{equation}
with primes indicating differentiation with respect to $\xi$.
An integration by parts, with the boundary term exactly 
canceled by the Gibbons-Hawking term, gives
\begin{equation} 
 S=2 \pi^2 \,\int_{\xi_{min}}^{\xi_{max}} d\xi \left\{ \rho^3 
  \left[ \frac{1}{2} \phi'^2 + U(\phi) \right] 
  - \frac{3}{\kappa} ( \rho \rho'^2 + \rho)\right\} \,.
\label{action2}
\end{equation}

The equations of motion that follow from this action are
\begin{eqnarray}
  \phi'' +3  \frac{\rho'}{\rho} \phi' & = & \frac{dU}{d \phi} \, ,
 \label{phieq} \\ 
 \rho'^2 &=& 1 + \frac{\kappa}{3} \rho^2 \left[  \frac{1}{2} \phi'^2- U(\phi) \right] \,.
 \label{rhoeq}
\end{eqnarray}

A further useful equation, obtained by differentiating Eq.~(\ref{rhoeq}) and then 
using Eq.~(\ref{phieq}), is
\begin{equation}
  \rho''= -\frac{\kappa}{3}\,\left[ \phi'^2 +U(\phi) \right] \rho \,.
\label{rhoPrimePrime}
\end{equation}

The boundary conditions for these equations depend on the topology of
the solution.  To avoid a singularity, $\phi'$ must vanish at any zero
of $\rho$.  In all bounce solutions there is at least one such zero,
which can be taken to lie at $\xi=0$.  Bounces describing decays from
a de Sitter vacuum have a second zero (and thus an $S^4$ topology),
but for decays from an anti-de Sitter or Minkowski vacuum, such as we
are concerned with in this work, any acceptable bounce must have an $R^4$ 
topology and thus only a
single zero of $\rho$.  We thus have
$\xi_{min}=0$ and $\xi_{max} =\infty$ and the boundary conditions
\begin{equation}
  \phi'(0)=0, ~~~\phi(\infty)=\phi_{\rm{fv}}, ~~~\rho(0)=0 \,.
\label{bdy-cond}
\end{equation}

The field equations can be used to obtain simpler expressions for the action
of the bounce solution.  First, as pointed out by CDL, substitution
of Eq.~(\ref{rhoeq}) into Eq.~(\ref{action2}) gives
\begin{equation}
  S=4\pi^2 \,\int_0^\infty d\xi 
  \,\, \left[\rho^3 U(\phi)-\frac{3}{\kappa} \rho \right] \,.
\end{equation}
Note that this integral will be divergent for both the bounce and for
the initial false vacuum.  
An alternative expression is obtained by substituting Eqs.~(\ref{rhoeq}) 
and (\ref{rhoPrimePrime}) 
for $\rho'^2$ and $\rho''$ into Eq.~(\ref{action1}).  This gives
\begin{equation}
   S = -2\pi^2 \,\int_0^\infty d\xi \,\,\rho^3 U(\phi)
   \, + \,\text {boundary~terms}  \,.
\end{equation}
For decay from a Minkowski vacuum the boundary terms of the bounce and
of the false vacuum cancel in the calculation of $B$.  Since $U(\phi)$
is identically zero in the false vacuum, we then have
\begin{equation}
  B= -2\pi^2 \,\int_0^\infty d\xi \,\,\rho^3 U(\phi)  \,.
\label{newBequation}
\end{equation}

\subsection{Thin-Wall Approximation}

Coleman and De Luccia computed the bounce action in the thin-wall
approximation. In this limit we can easily distinguish three regions
in the profile of the field $\phi$: a region of pure true vacuum
(inside the bounce), a region of pure false vacuum (outside the
bounce) and a well defined thin wall at $\rho=\bar\rho$ separating the
two regions. This approximation requires that the
energy difference between the two vacua,
\begin{equation}
  \epsilon= U(\phi_{\rm{fv}})- U(\phi_{\rm{tv}}) \, ,
\end{equation}
be small compared to the other mass scales in the potential. 

The existence of these distinct regions suggests splitting the
integration region in the actions into three parts, whose contributions
to the tunneling exponent are
\begin{eqnarray}
    B_{\rm{outside}} &=& 0 \, , \\
    B_{wall} & = & 2 \pi^2 \bar{\rho}^3 \sigma 
 \equiv 4 \pi^2 \bar{\rho}^3 \int d\xi ~[U(\phi)-U(\phi_{\rm{fv}}) ] \, ,\\
   B_{\rm{inside}}&= &\frac{12 \pi^2}{\kappa^2}  
 \left\{ \frac{1}{U(\phi_{\rm{tv}})}
    \left[ \left(1-\frac{\kappa}{3} \bar{\rho}^2 U(\phi_{\rm{tv}}) \right)^{3/2}
      -1\right] -(\phi_{\rm{tv}} \rightarrow \phi_{\rm{fv}} )\right\} \, .
\end{eqnarray}
These expressions only depend on the location $\bar\rho$ of the wall, the values
of the potential at the vacua, and the surface tension $\sigma$ of the wall.
The location of the wall is determined by requiring that it be 
a stationary point of the action.

For transitions from Minkowski to anti-de Sitter the wall is
located at
\begin{equation} 
   \bar{\rho}=\frac{\bar{\rho}_0}{1- (\bar{\rho}_0/2 \ell)^2} \, ,
\end{equation}
where $\bar{\rho}_0= 3 \sigma/\epsilon$ is the location of the wall in
the absence of gravity, and 
\begin{equation}
     \ell= (\kappa \epsilon/3)^{-1/2}
\end{equation}
is the anti-de Sitter radius. The tunneling exponent is
\begin{equation} 
      B=\frac{B_0}{[1- (\bar{\rho}_0/2 \ell)^2]^2} \, ,
\end{equation} where the
subscript zero indicates the value of the action in the absence of
gravitational effects.
If we define a critical value
\begin{equation}
    \kappa_{\rm cr} = \frac{4 \epsilon}{3\sigma^2} \, ,
\end{equation}
the above results can written as
\begin{equation}
  \bar\rho = \frac{\bar\rho_0}{\left({1 - \kappa/\kappa_{\rm cr}}\right)}
           \,  \, ,
\label{rhoDiverge}
\end{equation}
and 
\begin{equation} 
     B = \frac{B_0}{\left({1 - \kappa/\kappa_{\rm cr}}\right)^2} \, .
\label{Bdiverge}
\end{equation}
Gravity quenches vacuum decay when $\bar{\rho}_0 \ge 2 \ell$ or, equivalently,
when $\kappa \ge \kappa_{\rm cr}$.

For comparison with later results, we note that the thin-wall
approximation for a theory with a quartic potential with almost
degenerate vacua at $\phi=0$ and $\phi =v$ predicts that the field profile in
the wall is
\begin{equation}
     \phi = \frac{v}{2}\left\{\tanh\left[ 
  \frac{\mu}{2}\, (\xi - \bar \xi) \right] +1 \right\}\, ,
\end{equation}
where $\mu^2 = U''(\phi_{\rm fv})$.

\subsection{Energy considerations}

Quantum tunneling in Minkowski space must conserve energy.  With
gravitational effects ignored, this implies that the energy of the
spherically symmetric and instantaneously static configuration at the
time of bubble nucleation must vanish; i.e., 
\begin{equation}
    0 =  4\pi \int \, dr\,r^2\left(\frac12 \phi'^2 +U\right) \, .
\end{equation}
This constraint is modified a bit when gravity is taken into account.
Recall that a static spherically symmetric metric can be written in
the form
\begin{equation}
    ds^2 = -B(r) dt^2 + A(r) dr^2 + r^2 d\Omega_2^2 \, .
\end{equation}
Einstein's equations lead to
\begin{equation}
     A(r) = \left[1 - \frac{2G_N {\cal M}(r)}{r} \right]^{-1}  \, ,
\end{equation}
where
\begin{equation}
      {\cal M}(r) =4\pi \int_0^r ds \,s^2 \tilde\rho(s)
\end{equation}
with $\tilde\rho$ the energy density.  Conservation of energy becomes the 
requirement that the ADM mass $M={\cal M}(\infty)$ vanish.  
For our scalar field theory,
\begin{eqnarray}
    \tilde\rho &=& \frac{1}{2A} \, \left( \frac{d\phi}{dr}\right)^2 + U \cr
      &=& \frac12 \left( \frac{d\phi}{d\xi}\right)^2 + U \, .
\end{eqnarray}
If we identify $\rho$ with $r$ on the $x_4=0$ spacelike surface, 
our requirement becomes
\begin{eqnarray}
     0 &=& 4\pi \int_0^\infty d\rho \, \rho^2 \left(\frac12 \phi'^2 + U \right) \cr
     &=& 4\pi \int_0^\infty d\xi \, \rho^2 \rho' \left(\frac12 \phi'^2 + U \right) \, .
\label{Energy}
\end{eqnarray}

Now note that 
\begin{eqnarray}
   \frac{d}{d\xi}\left[-\rho^3\left(\frac12 \phi'^2 - U\right) \right]
     &=& \rho^3\left(-\phi' \phi'' +\phi' \frac{dU}{d\phi} \right)
        -3 \rho^2 \rho'\left(\frac12 \phi'^2 -U\right) \cr
     &=& 3 \rho^2 \rho' \left(\frac12 \phi'^2 +U\right) \, ,
\end{eqnarray}
where the second equality is obtained by using Eq.~(\ref{phieq}) to
eliminate $\phi''$. The quantity in brackets clearly vanishes at
$\xi=0$.  It also vanishes as $\xi \to \infty$, provided that $\phi'^2$
and $U$ fall faster than $1/\rho^3$, which they do for a bounce
describing decay from a Minkowski vacuum.

Hence, any solution of the Minkowski to AdS bounce equations is
guaranteed to conserve energy.  If energy cannot be conserved, then
tunneling is impossible. Indeed, both CDL and
S.~Weinberg~\cite{Weinberg:1982id} explain the absence of tunneling
when $\kappa \ge \kappa_{\rm cr}$ by showing that in this regime it is
impossible to construct a thin-wall bounce that conserves energy.  We
will see presently how this line of argument can be applied beyond the
usual thin-wall approximation.

\section{Thick-wall bounces --- Numerical results}
\label{thickwall-sec}

The thin-wall approximation has the attractive feature of producing
analytic expressions for the bounce action and for the field profile
of the bounce.  Unfortunately, its validity is restricted to a rather
special class of potentials, those for which $\epsilon$, the
difference in energy densities between the true and false vacua, is
small compared to the scale set by the potential barrier separating
the two vacua.  It is well known that in the absence of gravity a
generic potential that does not satisfy this criterion can yield a
``thick-wall'' bounce solution in which the scalar field, even at the
center of the bounce, is never close to its true vacuum value.  We
have explored a number of such potentials, following the evolution of
the bounce as the strength of gravity is increased.  In this section
we will use several examples to illustrate the generic features that
we have found.

First, we consider a number of quartic potentials.  An arbitrary
quartic potential can be written in the form
\begin{equation}
    U(\phi) = \lambda \left(C_0 + C_1 \phi
              + C_2 \phi^2 + C_3 \phi^3 + \phi^4
             \right) \, .
\label{quartic-phi-expression}
\end{equation}
We will assume that $\lambda$ is positive and that the other coefficients
are such that $U$ has two minima.

Variation of the constants gives five degrees of freedom.  Two,
corresponding to translations in the $U$-$\phi$ plane, are fixed by
our convention that the true vacuum be at $\phi_{\rm tv} =0$ and the
requirement that $U_{\rm fv}=0$; the former implies that $C_1=0$,
while the latter imposes a constraint relating $\epsilon=-\lambda
C_0$, $C_2$, and $C_3$.

Two more degrees of freedom correspond to rescalings of the height and
width of the potential.  The first of these is implemented by varying
$\lambda$.  At the classical level, the effects of this can be
absorbed by a rescaling of the length scale.  Simple scaling arguments
show that the bounce action is proportional to $1/\lambda$, while the
width of the field profile is proportional to $1/\sqrt{\lambda}$.  In
our investigation of the bounce solutions, we find it convenient to
work with $\lambda=1$, keeping in mind the possibility of restoring
$\lambda$ by a rescaling.  In particular, we will be led to consider
solutions with fields comparable to or greater than the Planck mass.
By choosing a suitably small value of $\lambda$, the energies
associated with these solutions can always be made sub-Planckian.

\begin{figure}
   \centering
   \includegraphics[width=3.0in]{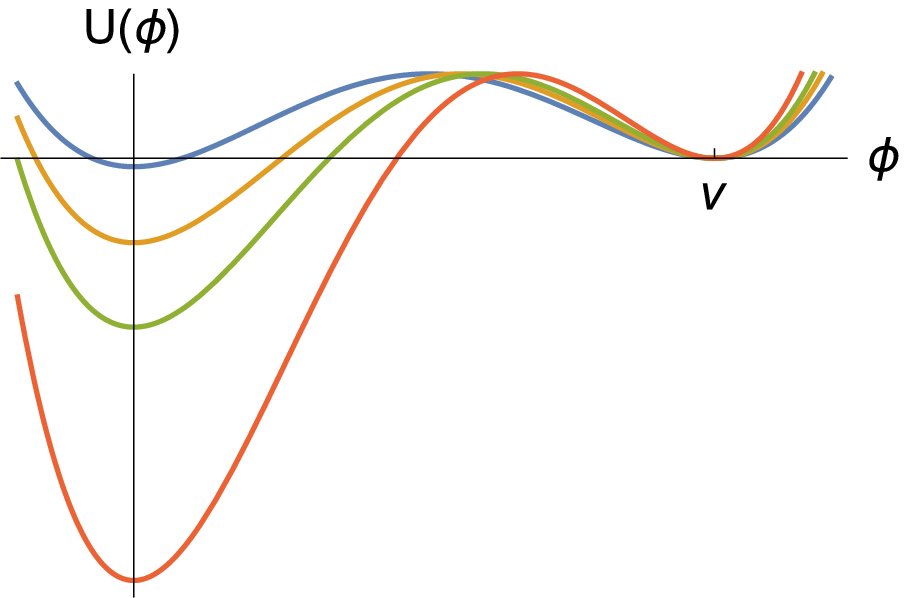}
   \includegraphics[width=3.0in]{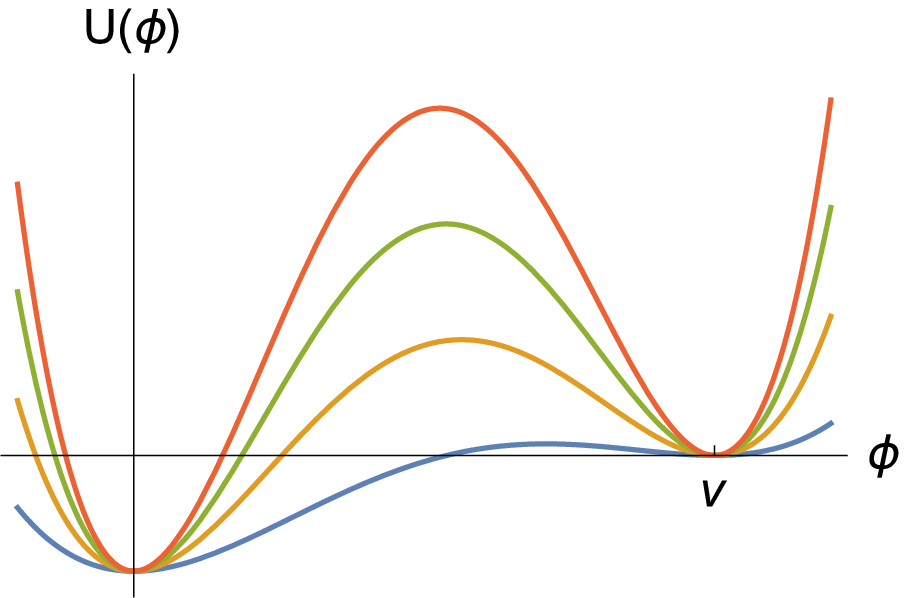}
   \caption{Quartic potentials with fixed $U_{\rm top}$ and 
   $\epsilon/U_{\rm top} = 0.1$, 1, 2, and 5 (on left)
 and fixed $\epsilon$ and $U_{\rm top}/\epsilon = 0.1$, 1, 2, and
  3 (on right).}
  \label{varyUtop}
\end{figure}

Rescaling the width of the potential is implemented by varying the
position of the false vacuum, which we denote by $\phi_{\rm fv} = v$.
This leaves one remaining degree of freedom.  Although this could be
implemented by varying a single coefficient in the potential, we find
it more instructive to parameterize it by the value of $U_{\rm top}$,
a quantity that can be naturally generalized to other, non-polynomial,
potentials.  In Fig.~\ref{varyUtop}, we show the effects of separate
variations of the dimensionless quantities $U_{\rm top}/v^4$ and
$\epsilon/v^4$.

We will investigate the effects of gravity by following the evolution
of the bounce solutions as the mass scales in the potential  are
increased.  In order to separate the effects of a stronger
gravitational interaction from those due to a variation of the field
dynamics, we will keep the shape of the potential unchanged.  Thus, we
will vary $v$ but hold $\epsilon/v^4$ and $U_{\rm top}/v^4$ 
fixed.  The quantity 
\begin{equation}
    \beta = \sqrt{\kappa v^2} =   \frac{\sqrt{8\pi} \,v}{M_{\rm Pl} }
\end{equation}
can then be viewed as a measure of the relative strength of the
gravitational interactions.  Thus $\beta$ and $U_{\rm top}/\epsilon$
can be taken as the dimensionless quantities spanning the 
parameter space of our potentials.

It may be helpful to relate our parameterization of the quartic
potential with that used in Refs.~\cite{Bousso:2006am} and
\cite{Aguirre:2006ap}.  For the case of a Minkowski false vacuum, that
potential can be written as
\begin{equation}
    U(\tilde\phi) = \left(\frac{\mu}{M}\right)^4 \left(-\frac12 M^2\tilde\phi^2
     - \frac{b}{3} M \tilde\phi^3 + \frac14 \tilde\phi^4\right)  + C
\end{equation}
where the constant $C$ is chosen so that $U_{\rm fv}=0$, and tildes have
been inserted because their field is shifted relative to ours, with the 
top of the barrier lying at $\tilde\phi=0$.  The strength of gravity 
is measured by the quantity
\begin{equation}
   \tilde\epsilon = \sqrt{\frac{8\pi}{3}}\, \frac{M}{M_{\rm Pl}}
\end{equation}
where we have inserted a tilde on their $\epsilon$ to avoid confusion
with our notation for the true vacuum energy density.

Variation of their $b$ corresponds to a variation of both
$\epsilon/v^4$ and $U_{\rm top}/v^4$ in our potential.  When $|b|$ is
small, the true and false vacuum are almost degenerate, and $U_{\rm
  top}/\epsilon$ is small; a large value of $|b|$ corresponds to a
large $U_{\rm top}/\epsilon$.  Although simple closed form expressions
converting between the two sets of conventions cannot be obtained in
general, matters simplify for the case $b=1$ that was used for the
numerical analysis in Refs.~\cite{Bousso:2006am} and
\cite{Aguirre:2006ap}.

With $b=1$, the true and false vacua and the top of the potential
barrier occur at 
\begin{eqnarray}  
     \tilde\phi_{\rm tv} &=& \frac{1+\sqrt{5}}{2}\, M   \cr
     \tilde\phi_{\rm fv} &=& \frac{1-\sqrt{5}}{2}\, M   \cr
     \tilde \phi_{\rm top} &=& 0 \, .
\end{eqnarray}
From this we find that the ratio of the barrier height to 
the true vacuum energy density is given by
\begin{equation} 
     \frac{ U_{\rm top}}{\epsilon} 
       = \frac{-25 + 13 \sqrt{5}}{50}
       = 0.08138  \, .
\label{their-value}
\end{equation}
The distance between the minima, $v$ in our notation, is then
\begin{equation}
      v = |\phi_{\rm fv} - \phi_{\rm tv}| = \sqrt{5} M
\end{equation}
so that our measure of the strength of gravity is related
to theirs by
\begin{equation}
      \beta = \frac{\sqrt{3}\,v}{M} \, \tilde\epsilon 
         = \sqrt{15}\,\,\tilde \epsilon \, .
\end{equation}

\begin{figure}
   \centering
   \includegraphics[width=3.2in]{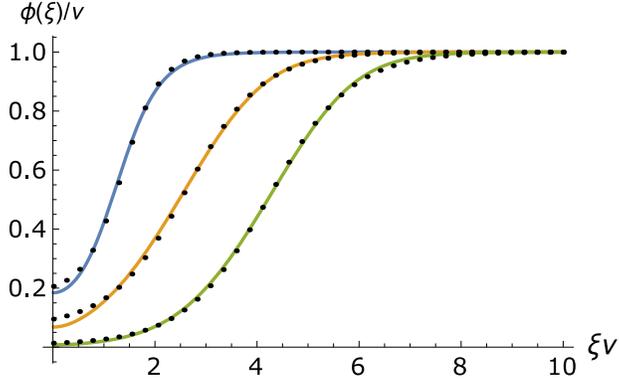}
   \caption{Profiles of $\phi$ for the quartic potential with $U_{\rm
       top}/v^4= 0.01 $ and $\epsilon/v^4=
     1$.  The solid curves indicate the best fit of a hyperbolic
     tangent to the data.  Reading from left to right, the curves
     correspond to $\beta = 0$, 5.747, and 5.763.}

   \label{quartic-fig1}
\end{figure}

\begin{figure}
   \centering
   \includegraphics[width=3.2in]{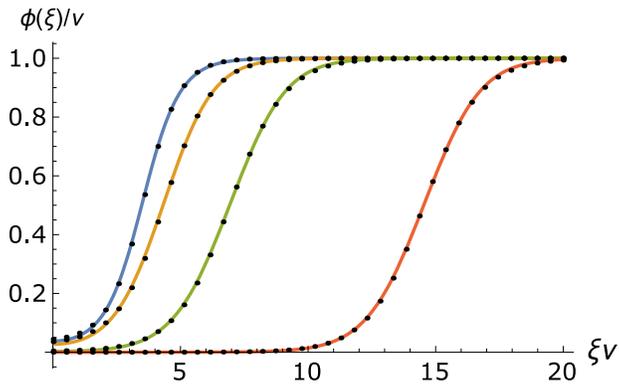}
   \caption{Similar to Fig.~\ref{quartic-fig1}, but for $U_{\rm top}/v^4 =
     0.01 $ and $\epsilon/v^4=
     0.2$. Reading from left to right, the curves
     correspond to $\beta = 0$, 2.70, 3.33, and 3.38.}
   \label{quartic-fig2} 
\end{figure}

\begin{figure}
   \centering
   \includegraphics[width=3.2in]{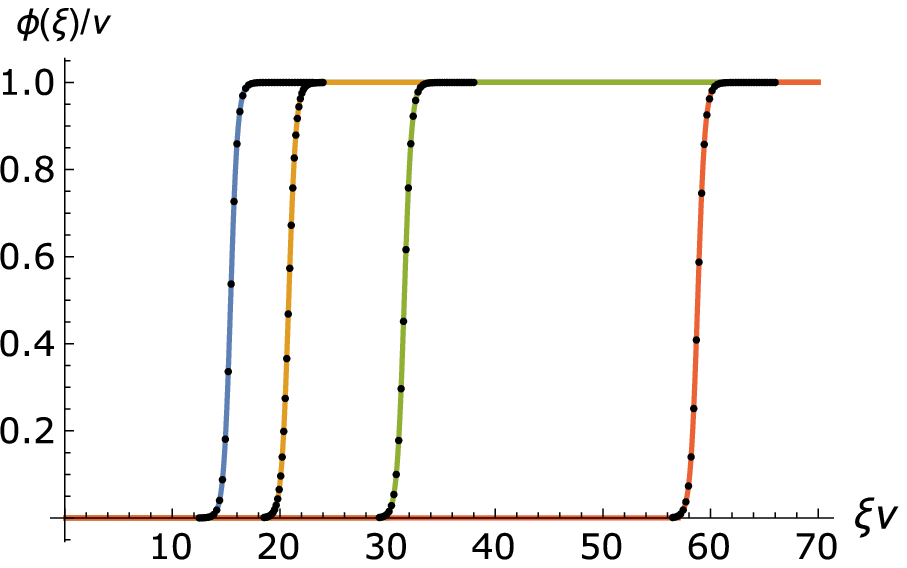}
   \caption{Similar to Fig.~\ref{quartic-fig1}, but for $U_{\rm top}/v^4 =
     0.25 $ and $\epsilon/v^4= 0.1$.  Reading from left to right, the
     curves correspond to $\beta = 0$, 0.588, 0.714, and 0.739.}
   \label{quartic-fig3}
\end{figure}

Figures~\ref{quartic-fig1}--\ref{quartic-fig3}
show the evolution of the field profile with $\beta$ for several
different choices of $\epsilon$ and $U_{\rm top}$.  Figure~\ref{quartic-fig1}
shows bounces for a potential with $\epsilon/U_{\rm top} = 100$.
In the weak
gravity limit, $\beta=0$, the bounce is
clearly a thick-wall solution, with the initial value of the field not
even close to $\phi_{\rm tv}$.  The corresponding bounce for a 
potential with $\epsilon/U_{\rm top} = 20$, shown in 
Fig.~\ref{quartic-fig2} begins closer to the true vacuum, but
immediately evolves rapidly toward $\phi_{\rm fv}$; although there is
a suggestion of a bubble wall, there is no clearly defined true vacuum
interior.  The bounces in Fig.~\ref{quartic-fig3} are closer to one's
conception of a thin-wall bounce, but with this moderately small ratio
$\epsilon/U_{\rm top}=0.4$, the quantitative predictions of the 
thin-wall analysis are not borne out.

However, in all three cases, with larger $\beta$
the transition from $\phi_{\rm tv}$ to
$\phi_{\rm fv}$ occurs at larger values of $\xi$ and there is a more
clearly defined transition region centered about a value $\xi =
\xi_0$.  As $\beta$ is increased, this transition region moves further
outward, but with its shape remaining roughly constant.  There is a
critical value $\beta_{\rm cr}$ beyond which there is no bounce.  As
this value is approached, $\xi_0$ rapidly increases and, we will argue,
tends to infinity in the critical limit.

In fact, the field profile in the transition region can be approximated by
a hyperbolic tangent, with 
\begin{equation}
      \phi(\xi) \approx  \frac{v}{2} \left\{
      \tanh[b(\xi-\xi_0)] + 1 \right\}  \, .
\label{tanhform}
\end{equation}
As can be seen in the figures, the fit is
surprisingly good, although there is a slight mismatch where 
the exponential tails of the hyperbolic tangent overlap with those
of the true and false vacuum regions.
We do not have an analytic explanation for this
behavior, which we have found in all of the examples that we have
examined.  As we saw in Sec.~\ref{setup-section}, one also finds a
hyperbolic tangent in the thin-wall approximation for a quartic
potential.  However, in that case the parameter $b$ could be directly
read off from the potential and did not depend on $\beta$.  In our
more general case $b$ varies with $\beta$ (even in the quasi-thin-wall
solutions of Fig.~\ref{quartic-fig3}), decreasing as gravitational
effects become stronger, and cannot be read off directly from 
the parameters in the potential.  In the next section we will discuss
the computation of $b$ in the critical limit.

There is another, quite significant difference from the thin-wall
approximation.  In that case, the transition region of the hyperbolic
tangent clearly maps onto the potential barrier between the true and
false vacua; there is no ambiguity in referring to it as the bounce
wall.  Matters are not so clear in the more general case.  When
$\epsilon$ is large compared to the height of the potential barrier,
only the very last part of the transition region of the hyperbolic tangent 
maps onto the region
where the potential is higher than in the false vacuum; i.e., onto
the true barrier.  The rest corresponds to an AdS-like region with a
negative cosmological ``constant'' that decreases in magnitude as one
moves outward.  To emphasize this distinction, we will refer to the
entire transition region as a ``step'', and use of the term ``wall''
to only refer to the region with positive potential.  More
specifically, we will define the wall to be the region lying between
$\phi_{\rm left}$, the point on the true-vacuum side where
$U(\phi)=U_{\rm fv}=0$, and $\phi_{\rm right}$, the point on the 
false vacuum side where $\phi$ has gone 90\%  of the way from the 
top of the barrier to the false vacuum.

\begin{table}
\caption{\label{data1} 
Data for bounces with the quartic potential, used in
Fig.~\ref{quartic-fig1}, with $U_{\rm top}/v^4 =0.01$ and
$\epsilon/v^4=1$ and several values of $\beta$, which is a measure of
the strength of gravity.  For this potential the critical value of
$\beta$, at which the tunneling exponent $B$ diverges and tunneling is
quenched, is $\beta_{\rm cr}=5.7633$.  The radius of the bounce wall,
in coordinate and geometric units, is given by $\xi_{\rm wall}$ and
$\rho_{\rm wall}$, while $\Delta\xi_{\rm wall}$ and $\rho_{\Delta\rm                      
  wall}$ give its thickness.  These should be compared with the AdS
length $\ell_{\rm AdS}$.  The last three columns show $\eta$, a
measure of how much the value of the field at the center of the bounce
departs from the true vacuum value; the wall action $\sigma_{\rm                          
  wall}$; and $b$, which parameterizes the field profile in the wall.
These quantities are defined more precisely in the text.  All
dimensionful entries are to be understood to be in units of
appropriate powers of $v$.}
\begin{ruledtabular}
\begin{tabular}{|c|ccccccccc|}
  $\beta$ &$B$&$\xi_{\rm wall}$& $\rho_{\rm wall}$& $\ell_{\rm AdS}$&$\Delta \xi_{\rm wall}$&$\Delta \rho_{\rm wall}$ & $\eta$ &  $\sigma_{\rm wall}$ & $b$ \\ \hline 
  0 &4.37 &1.92 & 1.92 & $\infty$&1.35&1.350 &0.2181&0.0120 & 1.33 \cr % \\    \hline 
  1.00 &4.752& 1.95& 2.04& 1.732& 1.38& 1.391& 0.2247 & 0.01193 & 1.30 \cr    %\\   \hline
  1.67 & 5.523& 1.99& 2.25& 1.039& 1.42& 1.462& 0.2349&  0.01236 & 1.26 \cr  %\\  \hline
  2.50 &7.509& 2.06& 2.73& 0.693& 1.50& 1.615& 0.2499&  0.01314& 1.18 \cr %\\ \hline
  3.33 &12.1& 2.18& 3.61& 0.520& 1.60& 1.883& 0.2622&  0.01416 & 1.08 \cr %\\ \hline
  5.00 &104.9& 2.61& 11.5& 0.346& 1.87& 4.042& 0.2442&  0.01685 & 0.87 \cr%\\ \hline 
  5.56 & 1364 & 3.05& 42.6& 0.312& 1.98& 12.13& 0.1814&  0.01796 & 0.77 \cr%\\ \hline
  5.71 & 25060 & 3.50& 180& 0.303& 2.01& 47.85& 0.1171&  0.01830 & 0.71\cr%\\ \hline
  5.75 & 2.2$\times 10^5$& 3.84& 545& 0.301& 2.02& 142.2& 0.08054&  0.01837 & 0.67 \cr%\\ \hline
  5.76 & 6.3$\times 10^6$& 4.35& 2918 & 0.301& 2.03& 756.7& 0.04402&  0.01840 & 0.65  \cr%\\ \hline 
\end{tabular}
\end{ruledtabular}
\end{table}

\begin{table}
\caption{\label{data2} The same as for Table~\ref{data1}, but for the potential used 
in Fig.~\ref{quartic-fig2}, with $U_{\rm top}/v^4=0.01$ and 
$\epsilon/v^4=0.2$.  For this potential $\beta_{\rm cr}=3.38058$.}
\begin{ruledtabular}
\begin{tabular}{|c|ccccccccc|}
  $\beta$ &$B$&$\xi_{\rm wall}$& $\rho_{\rm wall}$& $\ell_{\rm AdS}$&$\Delta \xi_{\rm wall}$&$\Delta \rho_{\rm wall}$ &$\eta$ &  $\sigma_{\rm wall}$ & $b$\\ \hline 
  0& 54.2 &4.32 &  4.318 &$\infty$&2.39&2.39& 0.05013  & 0.0216 & 0.70 \cr
  0.100& 54.36 & 4.31  &4.323 & 38.7  & 2.397& 2.398& 0.05014&  0.0216&0.70 \cr
  1.000&67.02  &4.42&4.03 &3.87 &2.472& 2.563& 0.05053& 0.0224&0.68  \cr
  2.500    &309.7& 5.17& 11.5& 1.549& 2.857& 4.269& 0.04094&  0.0264 &0.56 \cr
  2.564  &355.2& 5.24& 12.36& 1.51&2.881& 4.483& 0.03955&  0.0266&0.56 \cr
  2.703 & 500.0& 5.41& 14.81& 1.43&2.934& 5.082& 0.03598&  0.0272&0.54 \cr 
  3.226& 8678 & 6.88& 63.9& 1.201&  3.165& 16.77& 0.01193&  0.0295 & 0.48\cr
  3.367 & 1.1  $\times10^6$ & 9.54& 725.2& 1.15& 3.236& 173.1& 9.9$\times10^{-4}$&  0.0302&0.46 \cr
  3.378 & 4.1 $\times 10^7$ & 11.6& 4427 & 1.146&3.242& 1048 &1.3$\times10^{-4}$&  0.0303 & 0.46 \cr 
  3.380 & 1.7 $\times10^8$& 12.4& 9084 & 1.146& 3.243& 2148 &6.0$\times10^{-5}$ & 0.0303&0.46 \cr
\end{tabular}
\end{ruledtabular}
\end{table}

\begin{table}
\caption{\label{data3}
  The same as for Table~\ref{data1}, but for the potential, used in Fig.~\ref{quartic-fig2},
with $U_{\rm top}/v^4=0.25$ and $\epsilon/v^4=0.1$. 
 For this potential $\beta_{\rm cr}=0.73986$.}
\begin{ruledtabular}
\begin{tabular}{|c|ccccccccc|}
  $\beta$ &$B$&$\xi_{\rm wall}$& $\rho_{\rm wall}$& $\ell_{\rm AdS}$&$\Delta \xi_{\rm wall}$&$\Delta \rho_{\rm wall}$&  $\eta $ &  $\sigma_{\rm wall}$  & $b$ \\ \hline 
   0  & 9294 & 15.5 & 15.5 & $\infty$ & 1.5 & 1.5 &$2.2\times 10^{-19}$ &  0.36& 1.54  \cr
  0.200 & 10820& 15.83& 16.7& 27.39& 1.503& 1.591& 7.3$\times 10^{-20}$  &  0.37 & 1.54  \cr
  0.333 & 14640& 16.6& 19.48& 16.43& 1.51& 1.797& 8.0$\times 10^{-21}$    &  0.37& 1.53 \cr
  0.500 &3.2$\times10^3$& 18.64& 28.73& 10.95& 1.525& 2.482& 2.1$\times 10^{-23}$    &  0.37 & 1.51 \cr 
  0.556 &4.9$\times10^4$& 19.87& 35.87& 9.859& 1.531& 3.008& 5.9$\times 10^{-25}$    &  0.38 &1.51 \cr 
  0.588 &6.8$\times10^4$& 20.85& 42.58& 9.311& 1.535& 3.503& 3.3$\times 10^{-26}$    &  0.38 & 1.50 \cr
  0.625 &1.1$\times10^5$& 22.33& 54.77& 8.764& 1.54& 4.402& 4.3$\times 10^{-28}$    &  0.38 & 1.50  \cr
  0.667& 2.6$\times10^5$& 24.92& 83.49& 8.216& 1.546& 6.518& 2.1$\times 10^{-31}$     & 0.38 & 1.49 \cr
  0.714&2.0$\times10^6$& 31.59& 230.7& 7.668& 1.553& 17.35& 5.7$\times 10^{-40}$  &  0.38 & 1.48 \cr 
  0.735& 6.4$\times10^7$& 43.67& 1279 & 7.449& 1.557& 94.58&1.4$\times 10^{-55}$   &  0.38 & 1.48 \cr 
  0.739 &1.0$\times10^{10}$& 58.85& 10120 & 7.416& 1.557& 746.3& 2.9$\times 10^{-75}$     & 0.38 & 1.48 \cr

\end{tabular}
\end{ruledtabular}
\end{table}

The details of the bounces for these three potentials are summarized
in Tables~\ref{data1}--\ref{data3}. 
In each table the mass scale, indicated
by $\beta$, increases as one moves down the table, while the shape of
the potential is held fixed.  All dimensionful quantities are quoted
in appropriate units of $v$; (e.g., the AdS length $\ell$ is
given as a multiple of $v^{-1}$).  The deviation from the true vacuum
of the field at the center of the bounce is measured by the parameter
\begin{equation}
       \eta \equiv \frac{\phi(0) -\phi_{\rm tv}}{\phi_{\rm fv} - \phi_{\rm tv}}
     = \frac{\phi(0)}{v}  \, .
\end{equation} 
The position of the bubble wall, denoted by $\xi_{\rm wall}$ and
$\rho_{\rm wall}$, is taken to be the point where $\phi=\phi_{\rm
  top}$, its value at the top of the barrier.  The wall thickness,
denoted by $\Delta\xi_{\rm wall}$ and $\Delta\rho_{\rm wall}$, is the
distance between the points corresponding to $\phi_{\rm left}$ and
$\phi_{\rm right}$, as defined above.  The quantity $b$ is the parameter
that appears in the argument of the hyperbolic tangent in Eq.~(\ref{tanhform}).
Finally, we have defined a surface tension $\sigma_{\rm wall}$ by
\begin{equation}
       \sigma_{\rm wall} = %\frac{1}{\rho_{\rm wall}^3}
    2\int_{\xi_{\rm left}}^{\xi_{\rm right}} d\xi \,[U(\phi) -U(\phi_{\rm fv})]\, .
\end{equation}
% If the variation of $\rho$ in traversing the wall were negligible,
% this would reduce to the formula for the surface tension in the
% absence of gravity and to the expression used by CDL.

Notice that $B$ is less than 5 for the first two entries in $\beta$ in Table~I.  With 
such small values for the tunneling exponent one should worry about the 
validity of the dilute gas approximation that underlies the bounce formalism.
However, it should be kept in mind that the values in the table correspond to 
$\lambda=1$; for weaker coupling $B$ would be increased by a factor of $1/\lambda$.

\begin{figure}
   \centering
   \includegraphics[width=3.2in]{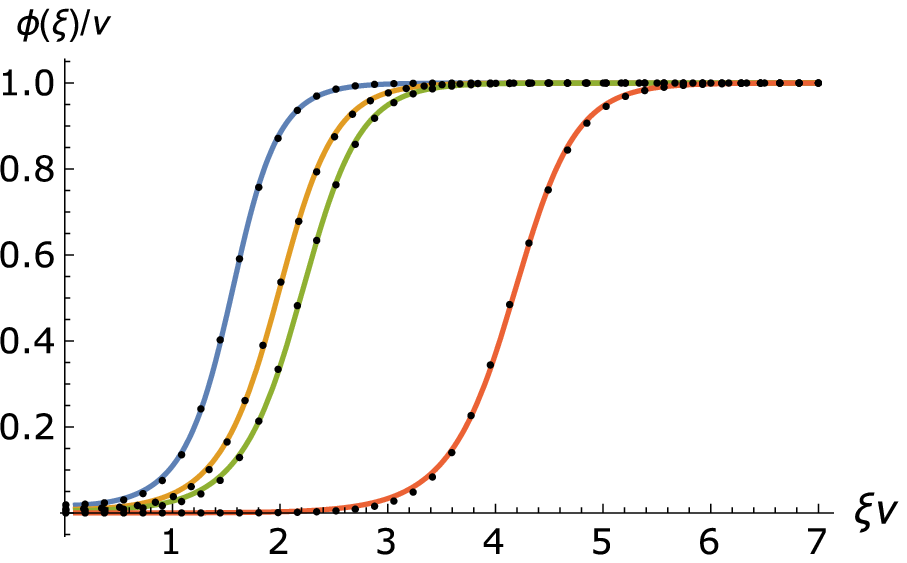}
   \caption{Similar to Fig.~\ref{quartic-fig1}, but for the potential of
     Eq.~(\ref{cosine-pot}), with $p=0.5$ and $\epsilon/U_{\rm top}= 7.30$.
     Reading from left to right, the curves correspond to $\beta =
     0$, 2.0, 2.2 and 2.5.}
   \label{cos-bounces}
\end{figure}

We have also considered a family of non-polynomial potentials of
the form
\begin{equation}
     U = \lambda v^4 \left[-\cos\left(\frac{\alpha \phi}{v}\right)
          - p \cos\left(\frac{3\alpha \phi}{v}\right)
           + q \right]
\label{cosine-pot}
\end{equation}
where $\alpha$ and $q$ are determined by the requirement that there be
a Minkowskian false vacuum at $\phi=v$.  As with the quartic
potentials, the effects of $\lambda$ at the classical level can be
absorbed by a rescaling, and so for convenience we set it equal to
unity.  As an example, in Fig.~\ref{cos-bounces} we show several
bounces for a potential with $p=0.5$, corresponding to $\epsilon/v^4 =
1.404$ and $U_{\rm top}/v^4 = 0.1925$, and give related data in
Table~\ref{cosine-data}.

\begin{table}
\caption{\label{cosine-data} The same as for Table~\ref{data1}, but for the potential
used in Fig.~\ref{cos-bounces}.  For this potential $\beta_{\rm cr}=2.5077$.}
\begin{ruledtabular}
\begin{tabular}{|c|ccccccccc|}
  $\beta$ &$B$&$\xi_{\rm wall}$& $\rho_{\rm wall}$& $\ell_{\rm AdS}$&$\Delta \xi_{\rm wall}$&$\Delta \rho_{\rm wall}$ & $\eta$ &  $\sigma_{\rm wall}$ & $b$ \\ \hline
0 &14.0 &1.71 & 1.72 & $\infty$&1.46&1.5 &0.0257&  0.13&2.17\cr
1.00 & 20.3& 1.80& 2.12& 1.5& 1.42& 1.5& 0.0234&  0.14&2.07  \cr
2.00 &116& 2.18& 5.37& 0.73& 1.57& 2.3& 0.01215&  0.16&1.81 \cr
2.25&412& 2.49& 10.3& 0.65& 1.62& 3.4& 6.4$\times 10^{-3}$&  0.16&1.72 \cr
2.40 &2249 & 2.93& 24.4& 0.61& 1.66& 6.4& 2.4$\times 10^{-3}$&  0.17 &1.66 \cr
2.45 &7712 & 3.26& 45.4& 0.60& 1.68& 10.9& 1.1$\times 10^{-3}$&  0.17 &1.65 \cr
2.48&3.3$\times 10^4$& 3.67& 94.2& 0.59& 1.68& 21.4& 4.1$\times 10^{-4}$ & 0.17 &1.63 \cr
2.49 & 8.1$\times 10^4$& 3.92& 147& 0.59& 1.69& 32.7& 2.2 $\times 10^{-4}$&  0.17&1.63 \cr
2.50 & 4.2$\times 10^5$& 4.39& 337& 0.58& 1.69& 73.3& 6.8$\times 10^{-5}$ & 0.17&1.63\cr
2.50628 & 1.2$\times 10^7$ & 5.35& 1780& 0.58& 1.69& 382& 6.2$\times 10^{-6}$ &  0.17&1.63 \cr
\end{tabular}
\end{ruledtabular}
\end{table}

\begin{figure}
   \centering
   \includegraphics[width=3.2in]{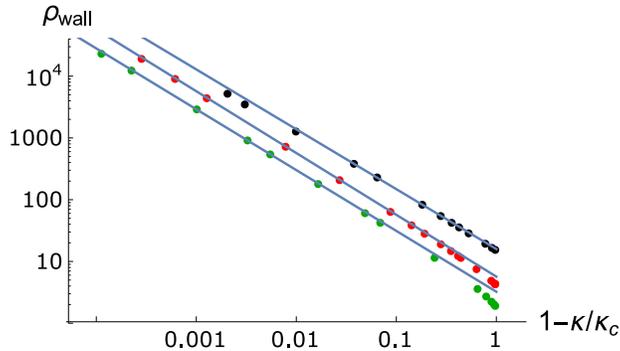}
   \caption{The behavior of $\rho_{\rm wall}$ (measured in units
  of $v^{-1}$) near the critical value of $\beta$.
     Reading from bottom to top, the green, red, and
     black circles represent the potentials for Tables~I, II, and
     III, respectively.  The best fit lines on this log-log plot have
     slopes -1.00, -0.99, and -0.97, respectively.  Note that in each
     case the last three points to the right have been omitted from
     the fit.}
   \label{rhoDivFig}
\end{figure}

\begin{figure}
   \centering
   \includegraphics[width=3.2in]{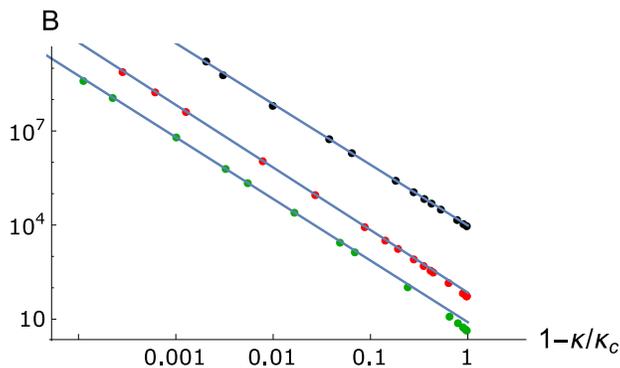}
   \caption{Similar to Fig.~\ref{rhoDivFig}, but for $B$.  The
slopes of the best fit lines are -2.00, -1.97, and -1.94. }
   \label{BDivFig}

\end{figure}

We have observed a number of regularities that 
persist in all of the examples we have considered:

1) Equations~(\ref{rhoDiverge}) and (\ref{Bdiverge}) show that in the
thin-wall limit both the bounce radius $\bar\rho$ and the tunneling
exponent $B$ diverge as $\kappa$ approaches a critical value, with
their values being proportional to $(\kappa_{\rm cr}-\kappa)^{-1}$ and
$(\kappa_{\rm cr}-\kappa)^{-2}$, respectively.
Figures~\ref{rhoDivFig} and \ref{BDivFig} illustrate similar behavior
for the potentials that we have examined.

There are, however, some differences.  In the thin-wall case the
inverse power dependence on $(\kappa_{\rm cr}-\kappa)^{-1}$ held for
the entire range of $\kappa$.  Here we see a deviation from this at
small values of $\kappa$; i.e., in the parameter range where the field
profile is the least ``thin-wall-like''.  We can quantify this by
comparing the the actual values of $\bar\rho_0$ and $B_0$ obtained
from our numerical solutions with $\kappa=\beta^2=0$ with the values
$\bar\rho_{\rm fit}$ and $B_{\rm fit}$ that would follow from
extrapolating the linear behavior all the way to $\kappa=0$. For the
potentials corresponding to Tables I, II, and III, we find 
$B_0/B_{\rm fit}=$0.55, 0.77, and 0.96 and 
$\bar\rho_0/\bar\rho_{\rm fit}=$0.59, 0.76, and 0.97.

There is another difference from the thin-wall case.  In that
limit, $\kappa_{\rm cr}$ corresponded to an AdS length that was
exactly half the bounce radius in the absence of gravity.  By
contrast, the limiting value of $\ell$ for our solutions is less than half of the
non-gravitational bounce radius.  This is consistent with the fact
that the average potential in the region contained within the bounce
wall is less than $\epsilon$ in magnitude, implying an effective AdS
length that is larger than that of the pure true vacuum.

2) The wall thickness as measured in $\xi$ initially increases
somewhat with $\beta$, but appears to tend to a finite value in the
critical limit.  By contrast, the thickness as
measured in $\rho$ diverges, with $\Delta\rho_{\rm
  wall}/\rho_{\rm wall}$ being roughly constant.

3) The surface tension $\sigma_{\rm wall}$ increases slightly with
$\beta$, but tends to a finite value in the critical limit.

4) Although $\eta$ may initially increase with $\beta$, as gravitational
effects continue to grow stronger $\eta$ decreases, eventually
vanishing in the critical limit.

\section{Analytic results}
\label{analytic-sec}

In this section we will describe how some understanding
of the numerical results of the previous section can be obtained.

The plots in Figs.~\ref{quartic-fig1}--\ref{quartic-fig3}
show three distinct
regions: Region I, $0<\xi<\xi_1$, where the field is exponentially
close to $\phi_{\rm tv}$, Region III, $\xi_3 <\xi < \infty$, where it
is exponentially close to $\phi_{\rm fv}$, and the step, Region II,
$\xi_1 < \xi < \xi_3$, that connects the other two.\footnote{There is 
some freedom in defining $\xi_1$ and $\xi_3$.  One could, e.g., 
arbitrarily choose them to be the values of $\xi$ at which $\phi$
is equal to $0.01v$ and $0.99v$.}
(As we have seen,
Region I is absent for some weak gravity solutions, but it always
appears once gravity is strong enough.)

In Region I $\phi$ is close to its true vacuum value, so to a good
approximation Eq.~(\ref{rhoeq}) is solved by
\begin{equation}
    \rho(\xi) = \ell \sinh(\xi/\ell) \, .
\label{rhoAsSinh}
\end{equation}
In the linearized approximation Eq.~(\ref{phieq}) then reduces to
\begin{equation}
    \phi'' + \frac{3\cosh(\xi/\ell)}{\ell \sinh(\xi/\ell)}\, \phi' 
     =U''(\phi_{\rm tv}) (\phi-\phi_{\rm tv})
           \equiv \mu_t^2 (\phi-\phi_{\rm tv}) \, .
\end{equation}
Changing
variables to $y=\cosh(\xi/\ell)$ and setting $\phi_{\rm tv}=0$ gives
\begin{equation}
   (y^2-1) \frac{d^2\phi}{dy^2} + 4y  \frac{d\phi}{dy} 
     = \ell^2 \mu_t^2 \phi  \, .
\end{equation}
This has the solution 
\begin{equation}
   \phi(\xi) =\phi(0)\,\frac{C_\alpha^{(3/2)}[\cosh(\xi/\ell)]}{C_\alpha^{(3/2)}(1)}
     = \phi(0) \, \frac{2C_\alpha^{(3/2)}[\cosh(\xi/\ell)]}{(\alpha+2)(\alpha+1)} \, ,
\end{equation}
where $C_\alpha^{(3/2)}(y)$ is a Gegenbauer function of the first
kind, with $\alpha(\alpha+3) = \mu_t^2 \ell^2$.  The fact that
$C_\alpha^{(3/2)}(1)$ is finite for all real positive $\alpha$
guarantees that $d\phi/d\xi$ vanishes at $\xi=0$, as required; it is
this boundary condition that eliminates the Gegenbauer function of the
second kind, $D_\alpha^{(3/2)}(y)$. 
If $\xi/\ell$ is moderately large, we can use the large argument approximation 
for the Gegenbauer function to obtain 
\begin{equation}
     \phi  \sim e^{a\xi/\ell}
\end{equation}
with 
\begin{equation}
   a = -\frac32 +\sqrt{\frac94 + \mu_t^2\ell^2} \, .
\end{equation}

In Region III $\phi$ is close to its false vacuum value, so $\rho$ 
is almost linear in $\xi$, with 
\begin{equation}
       \rho(\xi) \approx \rho(\xi_3) + (\xi -\xi_3) \, .
\end{equation}
The linearized equation for $\phi$ then becomes
\begin{equation}
   \phi'' + \frac{3}{\rho} \phi' = U''(\phi_{\rm fv}) (\phi-\phi_{\rm fv})
     \equiv \mu_f^2 (\phi-\phi_{\rm fv}) \, .
\end{equation}
As we approach the critical solution $\rho(\xi_3)$ 
becomes exponentially large, so the second term in this equation
can be ignored and we have
\begin{equation}  
     \phi_{\rm fv} -\phi(\xi) \sim e^{-\mu_f \xi} \, .
\end{equation}

To understand Region II, which connects these two regions,
it is helpful to follow Coleman's insight~\cite{Coleman:1977py} and view
the problem in terms of a particle
moving in an upside-down potential $-U(\phi)$ while subject to a
frictional force given by the $\phi'$ term in Eq.~(\ref{phieq}).    This
suggests defining a pseudo-energy
\begin{equation}
     E = \frac12 \phi'^2 - U(\phi) \, .
\end{equation}
This is not conserved; instead, the friction term in Eq.~(\ref{phieq})
causes it to decrease with $\xi$, with\footnote{The CDL formalism
  allows anti-friction, with $E'>0$; this is always encountered in
  decays from a de Sitter false vacuum, and might seem to be possible
  here when crossing the potential barrier where $U$ is positive.
  However, a bounce that has a period of anti-friction is necessarily
  compact, and therefore cannot describe decay from a Minkowski (or
  anti-de Sitter) vacuum.}
\begin{equation}
    E' = -3 \frac{\rho'}{\rho} \, \phi'^2 \, .
\end{equation}

\begin{figure}
   \centering
   \includegraphics[width=2.0in]{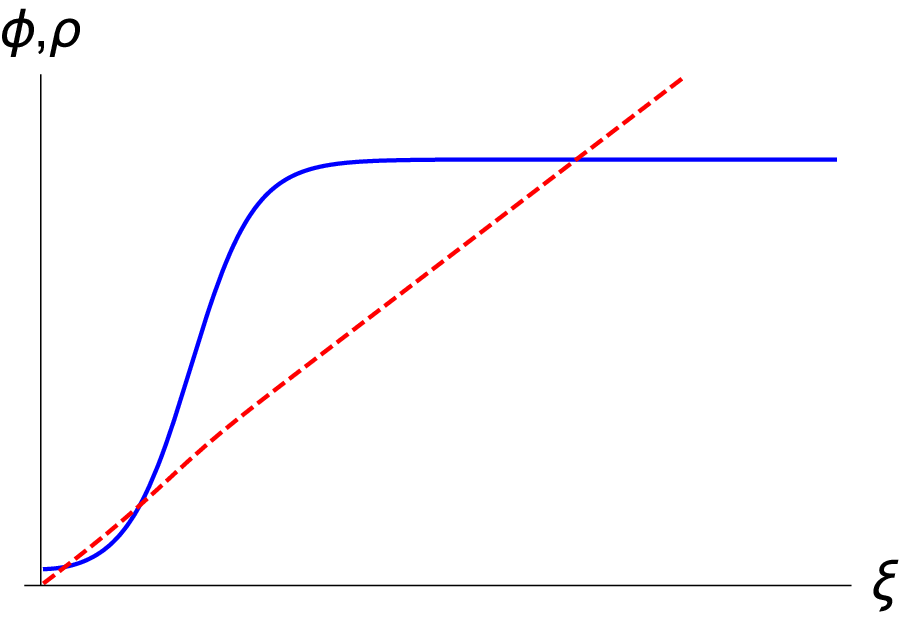}
   \includegraphics[width=2.0in]{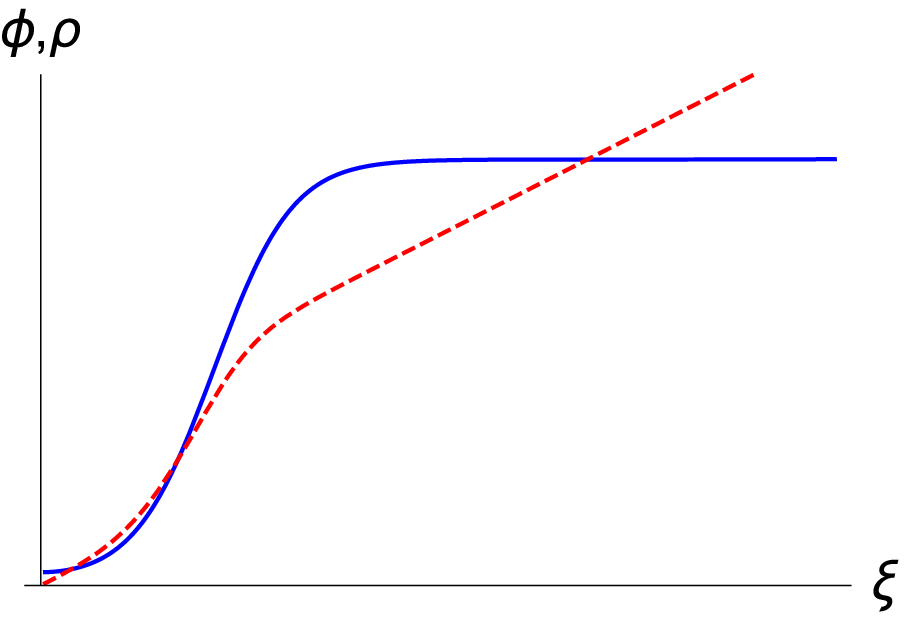}
   \includegraphics[width=2.0in]{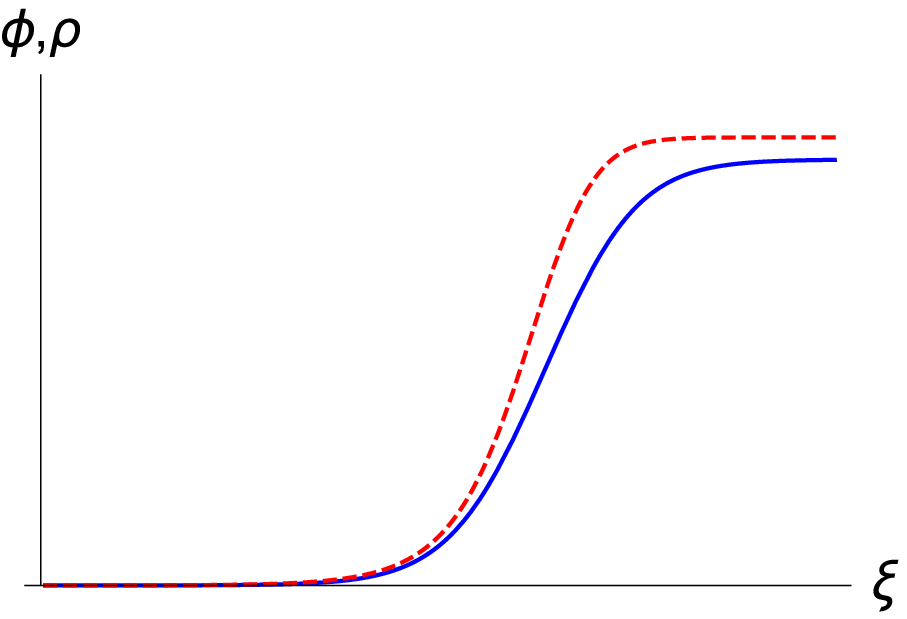}
   \caption{The metric function $\rho$ (dashed red line) and field
     $\phi$ (solid blue line) in three examples with, from left to
     right, weak, medium, and strong gravity.  The metric has been
     rescaled (by factors of 0.1, 0.005, and 0.00005, respectively) in
     order to fit it on the same graph as $\phi$.  In all three cases
     the actual asymptotic slope of $\rho$ is unity.  Note how the
     transition from Region~IIa to Region~IIb occurs later as the
     strength of gravity increases.}
   \label{RegionsIIa-IIb}
\end{figure}

Now note that Eq.~(\ref{rhoeq}) can be recast as 
\begin{equation}
    \frac{\rho'^2 }{\rho^2} = \frac{1}{\rho^2} + \frac{\kappa}{3}E \, .
\label{RhoPrimeRho}
\end{equation}
For all but the very beginning of Region I, the second term
dominates the right-hand side, and we can make the approximation
\begin{equation}
   \frac{\rho'}{\rho} = \sqrt\frac{\kappa}{3}\,\sqrt{E} \, .
\label{RhoPrimeApprox}
\end{equation}
This continues to hold in Region II until we reach a point $\xi_2$,
where $E$ has decreased enough that $(\kappa/3)E \approx
1/\rho^2$. We will refer to this interval, $\xi_1 <
\xi <\xi_2$, as Region IIa.  In Region IIb, $\xi_2 < \xi <\xi_3$,
the first term dominates the right hand side of Eq.~(\ref{RhoPrimeRho}).
(See Fig.~\ref{RegionsIIa-IIb}.)

In Region IIa, Eq.(\ref{RhoPrimeApprox}) leads to
\begin{equation}
   E' = - \sqrt{3 \kappa}\, \phi'^2 \sqrt{E} \, .
\label{Reg2Edot}
\end{equation}
Integrating this and noting that $E(\xi_1)\approx \epsilon$ leads to
\begin{equation}  
      \sqrt{E(\xi)} = \sqrt{\epsilon} 
     -\sqrt{\frac{3 \kappa}4} \int_{\xi_1}^\xi d\xi \, \phi'^2 \, ,  
\label{case-a-E}
\end{equation}
from which we obtain
\begin{equation}
   \rho(\xi) = \rho(\xi_1) \exp\left[\int_{\xi_1 }^\xi d\xi \,
       \sqrt{\frac{\kappa}{3}}\,\sqrt{E} \right] \, .
\label{RhoSolution}
\end{equation}

In Region IIb (and indeed also in Region III)
\begin{equation}
   \rho(\xi) \approx \rho(\xi_2) + (\xi - \xi_2)
\label{rhoLinear}
\end{equation}  
with the linear behavior indicating that the space is
approximately Minkowskian.  The pseudo-energy is
\begin{equation}
    E(\xi) = E(\xi_2)  - 3 \int_{\xi_2}^\xi d\xi \, 
     \frac{1}{\rho(\xi_2) + (\xi - \xi_2)} \, \phi'^2 \, .
\end{equation}

As the gravitational interactions grow stronger, the transition from
Region II to Region III takes place at ever larger values of $\xi_3$.
In the critical limit $\xi_3 \to \infty$, the Minkowskian false vacuum
is never reached and the bounce solution disappears.  This limit is
signaled by the vanishing of $E$ while Eq.~(\ref{case-a-E}) is still
valid.  This implies that
\begin{equation}
  \int_{\xi_1}^\infty d\xi \, \phi'^2 = \sqrt{\frac{4 \epsilon }{ 3\kappa}}
     \qquad \text{(critical~solution)} \, .
\label{phidotint}
\end{equation}

We can now understand why the field profile in Region~II remains 
relatively constant for large values of $\beta$ and $\kappa$, even
as the step itself moves to larger and larger $\xi$.
As the critical
value of $\beta$ is approached, the initial value of $\phi$ moves closer
to the true vacuum, with the result that the transition
from Region~I to Region~II occurs at a larger value of $\xi_1$.
Because $\rho$ grows exponentially with $\xi$ in region I [see Eq.~(\ref{rhoAsSinh})],
its value at the transition to Region~II, $\rho(\xi_1)$, increases 
exponentially with increasing $\beta$.

We next note that Eq.~(\ref{Reg2Edot}), which governs the field
profile in Region~IIa, can be rewritten as
\begin{equation}
    \phi'' + \sqrt{3\kappa} \,\sqrt{\frac12 \phi'^2 - U(\phi)}
        \,\,\, \phi' = \frac{\partial U}{\partial\phi}  \, .
\label{ddphiEq}
\end{equation}
Because $\rho$ does not appear in this equation, the
profile for $\phi$ as a function of $\xi$ does not depend on the value
of $\rho$ in Region~IIa.  Further, Eq.~(\ref{RhoSolution}) tells us that the 
growth factor $\rho(\xi)/\rho(\xi_1)$ is also independent of 
$\rho$ in this region.  

Now consider Region~IIb.  The transition from Region~IIa occurs
at $\xi_2$, the point where $(\kappa/2)E=1/\rho^2$.  As $\rho(\xi_1)$,
(and therefore $\rho$ throughout Region~IIa) increases, this transition
occurs at an ever larger value of $\xi$.  
This has several consequences.  First, Region~IIb accounts for a
smaller and smaller fraction of Region~II.  Second, 
with $\rho(\xi_2)$ exponentially
large in the near-critical regime, the friction term is negligible in
Region~IIb, implying a field profile that is independent of $\rho$.
Finally, we see from Eq.~(\ref{rhoLinear}) that the variation of $\rho$ 
in this region is negligible. 

The net result is that in the near-critical regime we can write
\begin{equation}
    \rho(\xi) = \rho(\xi_1) F(\xi-\xi_1)  \, ,
\end{equation}
where 
\begin{equation} 
       F(\xi-\xi_1) = \exp\left[\sqrt{\frac{\kappa}{3}} 
    \int_{\xi_1}^\xi d\xi \,\theta(\xi_2-\xi)\sqrt{E}\right] \, .
\end{equation}

In the thin-wall approximation, the absence of tunneling at large
$\kappa$ could be explained by the impossibility of finding a wall
radius $\bar\rho$ large enough that the negative contribution from the
true vacuum interior exactly canceled the positive wall contribution
to give zero net energy.  We can now see how this generalizes to the 
thick wall case.  Energy conservation requires that the energy integral in 
Eq.~(\ref{Energy}) vanish.  The contribution to this integral from Region~I
is 
\begin{equation}
     \int_0^{\xi_1} d\xi \, \rho^2 \rho' U = -\frac{\epsilon}{3} \rho(\xi_1)^3 \, , 
\label{TVcontrib}
\end{equation}
up to exponentially small corrections.  
The contribution from Region~II can be written, with the aid of 
Eqs.~(\ref{RhoPrimeApprox}) and (\ref{RhoSolution}), as
\begin{equation}
    \int_{\xi_1}^{\xi_2} d\xi\, \rho^3\, \frac{\rho'}{\rho}
           \left(\frac12 \phi'^2 + U \right) 
    = \rho(\xi_1)^3 \sqrt{\frac{\kappa}{3}}\int_{\xi_1}^{\xi_2} d\xi\,F(\xi-\xi_1)^3 
          \sqrt{E} \left(\frac12 \phi'^2 + U \right)  \, .
\label{StepContrib}
\end{equation}
In the strong gravity, near critical regime,
Region~IIa is almost all of Region~II, and so the total energy is well approximated
by the sum of Eqs.~(\ref{TVcontrib}) and (\ref{StepContrib}).
(The contribution from Region~III is exponentially small and can be neglected.)

The crucial point to note is that $\xi_1$ only enters through the
factors of $\rho(\xi_1)^3$ that appear in both contributions.  Because
these factors are common to the contributions of Regions~I and II, the
flat space or weak gravity strategy of arranging a cancellation
between these two by adjusting the size of the true vacuum region
fails, energy cannot be conserved, and the bounce solution disappears.

We now address the determination of $\beta_{\rm cr}$.  The most obvious 
method for doing this is to obtain bounce equations for many values of 
$\beta$, look for the rapid growth in $B$ and $\rho_{\rm wall}$ that is
seen in Tables~\ref{data1}-\ref{cosine-data}, and then extrapolate
to find the value of $\beta$ where these diverge.  

A second, computationally simpler, approach utilizes Eq.~(\ref{ddphiEq}).
A bounce solution will satisfy this equation throughout Region IIa,
where the pseudo-energy term dominates Eq.~(\ref{rhoeq}), but 
using this equation in Region IIb 
would
understate the friction there. A moment's thought
suffices to show that if a bounce exists, then using 
Eq.~(\ref{ddphiEq}) throughout Region II would give solutions
that start at rest on the true
vacuum side of the barrier and either reach or overshoot the false
vacuum.  In the limit of weak gravity, with $\kappa$ infinitesimal,
the friction term is negligible, and conservation of pseudo-energy
implies that overshooting solutions of this type will exist.  On the
other hand, if $\kappa$ is sufficiently large, the friction term will
be strong enough to prevent $\phi$ from ever reaching the false
vacuum.  The critical value of $\beta$ corresponds to the boundary
between these two regimes.  

An even simpler method utilizes our finding that the bounce is well
approximated by the hyperbolic tangent form of Eq.~(\ref{tanhform}).
Doing so and, with little error, setting the lower limit of the
integral in Eq.~(\ref{phidotint}) equal to $-\infty$, we obtain
\begin{equation}
     \frac13  bv^2
     = \sqrt{ \frac{4\epsilon}{3\kappa_{\rm cr}}}
\end{equation}
or
\begin{equation}
    \beta_{\rm cr}  = \frac{\sqrt{12 \epsilon}}{bv} \, .
\label{betaFROMb}
\end{equation}

The data in Table~\ref{beta-table} show that the three methods of
determining the critical value of $\beta$ are in good agreement.

\begin{table}
\caption{\label{beta-table} The values of $\beta_{\rm cr}$ for 
the potentials considered in Tables~\ref{data1}-\ref{cosine-data} obtained
by three different methods: (i) extrapolating from the data in the tables; (ii) 
shooting, using Eq.~(\ref{ddphiEq}); and (iii) inserting the measured value of 
$b$ into Eq.~(\ref{betaFROMb}).   }
\begin{ruledtabular}
\begin{tabular}{|c|ccc|}
& Extrapolation & Shooting & Using $b$ \\ \hline
Table I  & 5.76 & 5.75 & 5.75 \cr
Table II & 3.38 & 3.38 & 3.37 \cr
Table III & 0.739 & 0.737& 0.735 \cr
Table IV & 2.50 & 2.50 & 2.52 \cr
\end{tabular}
\end{ruledtabular}
\end{table}

\section{Concluding remarks}
\label{conclusion}

In this work we have revisited quantum tunneling from a Minkowski
false vacuum to an anti-de Sitter true vacuum.  We have examined
polynomial and non-polynomial potentials and scanned their parameter
spaces. We have found, through a combination of analytical and
numerical methods, that there is always a region of parameter space in
which tunneling is quenched by gravity; i.e., for any given potential,
increasing the effects of gravity by uniformly raising the mass scales
in that potential will eventually render the false vacuum stable.
This confirms and extends previous
work~\cite{Bousso:2006am,Aguirre:2006ap, Samuel:1991dy}.

\begin{figure}
   \centering
\includegraphics[width=3.2in]{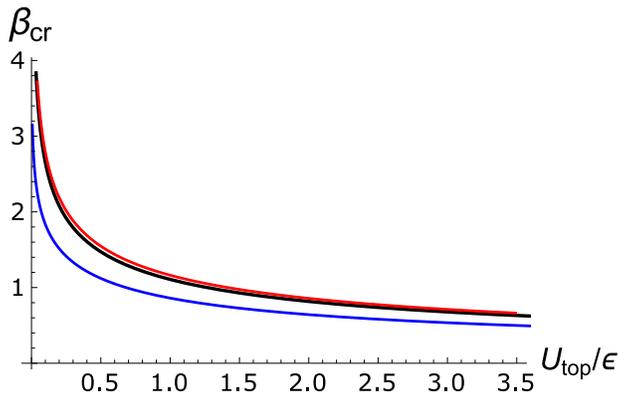}
   \caption{ Boundary of the stability region.  Decay from the
     Minkowski false vacuum is completely quenched in the region of
     parameters space above and to the right of the curves. The black
     curve corresponds to the quartic potential, the red curve just
     barely above it to the non-polynomial potentials of
     Eq.~(\ref{cosine-pot}), and the blue to a family of potentials
such as that in Fig.~\ref{weird-pot}. } 
   \label{criticalLine}
\end{figure}

As discussed in Sec.~\ref{thickwall-sec}, the relevant parameter space
for our potentials can be spanned by two dimensionless couplings,
$\beta$ and $U_{\rm{top}}/\epsilon$.  The former is the distance in
field space between the true and false vacua divided by the Planck
mass, while the latter is the ratio of the barrier height to the depth of
the true vacuum.  Figure~\ref{criticalLine} shows the critical
line that separates the region where the false vacuum is stable
against decay by bubble nucleation from the region where nucleation is
possible.  To determine this line it is not necessary to solve the coupled
equations for $\phi$ and $\rho$.  Instead, $\beta_{\rm cr}$ can 
be obtained by the computationally simpler overshoot/undershoot analysis
of Eq.~(\ref{ddphiEq}) described in Sec.~\ref{analytic-sec}.
 
As an aside, we recall from Eq.~(\ref{their-value}) that that the
numerical analysis of Ref.~\cite{Aguirre:2006ap} corresponded to a
value $U_{\rm top}/\epsilon = 0.8138$. Our shooting method then gives
$\beta_{\rm cr}=2.846$, and hence $\tilde\epsilon=0.7346$, consistent
with their result $\tilde\epsilon \approx 0.74$.

\begin{figure}
   \centering
\includegraphics[width=3.2in]{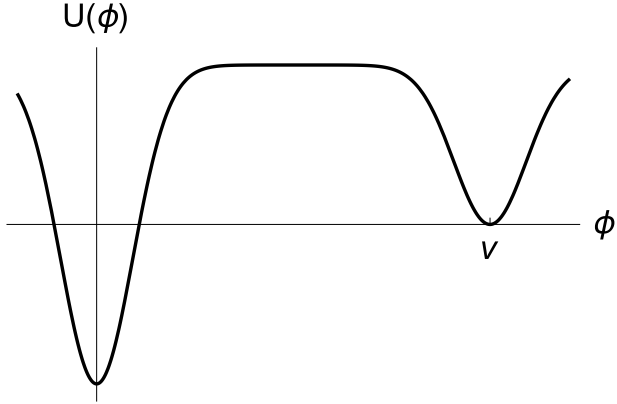}
   \caption{ One of the family of potentials, with two
     Gaussian potential wells, that corresponds to the blue line in
     Fig.~\ref{criticalLine}.}
   \label{weird-pot}
\end{figure}

Figure~\ref{criticalLine} shows the critical curves for
  several different families of potentials.  The black curve,
  corresponding to our quartic potentials, and the red curve, for the
  non-polynomial potentials of Eq.~(\ref{cosine-pot}), differ 
very slightly in the transition region, but have the same asymptotic 
behaviors.  The blue curve corresponds to a family of potentials of the 
form shown in Fig.~\ref{weird-pot}.
For all three cases the portion of the critical line extending to the
right, at large values of $U_{\rm{top}}/\epsilon$, agrees well with the
thin-wall approximation of CDL which predicts a power law
behavior $\beta_{\rm cr} \sim (U_{\rm top}/\epsilon)^{-1/2}$.
However, at smaller values of $U_{\rm{top}}/\epsilon$ the curves bend
upward, revealing a region where tunneling is possible.  The
continuations of the curves upwards to large values of $\beta$ can
also be fit by power laws.  For the black and red curves the power is
$-0.33$, while for the blue  curve the power is $-0.17$, a value
that varies as the width of the two wells is changed.

We expect the behavior displayed by these curves to emerge for all
classes of potentials.  The portion of the curve to the lower right,
where $\epsilon$ is small compared to the potential barrier,
corresponds to the familiar thin-wall bounces of CDL.  The portion on
the upper left, where the barrier becomes small, corresponds to a new
type of thin-wall bounce.  These evolve from thick-wall bounces as
$\beta_{\rm{cr}}$ is approached.  They are identical to the CDL-type
inside and outside, but differ in the structure of the shell that
separates these two regions. For the new bounces, the spherical region
of AdS vacuum is enclosed by a transition region composed of an inner
shell with a spatially varying negative energy density and an outer
shell where the matter field transverses the energy barrier. As
$\beta$ is increased the field profile and the thickness of this
region (measured in the appropriate coordinate) change little while
the radius of the true vacuum region increases and becomes infinite at
$\beta_{\rm{cr}}$.  These results can be generalized to the case of
transitions where the initial false vacuum is AdS.

The explanation of quenching in this new class of thin-wall bounces is
similar to that for the more familiar CDL class.  Tunneling is only
possible if an energy-conserving bubble configuration can be
constructed.  In flat spacetime this can always be done by making the
bubble radius large enough that the negative volume energy compensates
for the positive tension in the bubble wall.  Because of the peculiar
geometry of anti-de Sitter spacetime, this strategy begins to fail as
the bubble radius approaches the AdS length $\lambda$. This can occur
if a small $\epsilon$ makes the bubble large even in the absence of
gravity, but it can also happen if strong gravitational effects reduce
$\lambda$ relative to the length scales in the potential.

The results displayed in Fig.~\ref{criticalLine} imply that for a
given potential there is a set of parameters for which there are
stable Minkowski vacua and, by continuation, very long-lived de Sitter
vacua~\cite{Bousso:2006am, Aguirre:2006ap}.  We could be living in
such a vacuum, but our universe could instead be in a vacuum lying
below the critical line if that its bubble nucleation rate were
sufficiently small; i.e., if the tunneling exponent were sufficiently
large.  The minimum acceptable value of $B$ depends on the value of
the dimensionful prefactor $A$ in Eq.~(\ref{Gamma-form}).  If $A\sim
M_{\rm Pl}^4$, then our universe could have survived $\sim 10^{10}$
years if $B \gtrsim 600$; smaller values of $A$ would give smaller
bounds on $B$.  When comparing these bounds to the data in our
tables, it should be kept in mind that the overall coupling $\lambda$
multiplies the tabulated value of $B$ by $1/\lambda$.

In the literature one finds forms for the bounce action that differ in
the boundary terms.  This is not an issue for transitions from a de
Sitter vacuum, because these are governed by compact bounces.
However, this is not the case for decays from a Minkowski or anti-de
Sitter vacuum, where the corresponding bounces are non-compact.  It is
shown in Appendix~\ref{boundary-app} that with proper regularization
the tunneling exponent $B$ is finite and independent of whether or not
the Gibbons-Hawking boundary term is included.

Lastly, while supersymmetry, either exact or nearly so, is known to
suppress tunneling~\cite{Weinberg:1982id,Dine:2009tv}, the quenching
phenomenon occurs more generally; indeed, the specific potentials we
have considered are not supersymmetric.  It would be instructive to
apply the approach of this work to the special case of almost
supersymmetric theories in order to examine more closely the approach
to exact supersymmetry, and thus stability, and to understand how our
results generalize to theories with more fields and non-canonical
kinetic terms.

\begin{acknowledgments}

We are grateful to Ken Olum for helpful discussions and advice.
S.P. thanks the Institute for Advanced Study for hospitality and
support during the Spring of 2015 when this work was started.  She and
E.W. are grateful for the hospitality of the Aspen Center for Physics,
which is supported in part by the U.S. National Science Foundation
under Grant No.  PHYS-1066293.  This work was also supported in part
by the U.S. Department of Energy under Grant No. DE-SC0011941, by the
National Science Foundation under Grant Nos. PHY-1213888, PHY-1521186,
and PHY-1316033, and by a grant from the Simons Foundation (\#305975
to Sonia Paban).
\end{acknowledgments}

\appendix

\section{Boundary terms in the bounce action}
\label{boundary-app}

In many discussions of vacuum decay, including the original CDL paper,
the Gibbons-Hawking boundary term is omitted from the action.
In this appendix we show that while the 
inclusion or omission of this boundary term changes the actions 
of the bounce and of the initial false vacuum, the tunneling exponent
$B$, given by the difference between these two actions, is unchanged.
We also show that, although the individual actions may be divergent,
proper regularization gives a finite result for $B$ in both cases.
We will assume O(4) symmetry from the start.

We define two actions by 
\begin{eqnarray}
      S^{(1)} &=& 2\pi^2 \int d\xi\, \left[\rho^3 \left(\frac12 {\phi'}^2 + U \right)
        + \frac{3}{\kappa} \left( \rho^2\rho'' + \rho{\rho'}^2 - \rho
    \right)\right] \cr
         &=& -2 \pi^2 \int d\xi \, \rho^3 U 
\end{eqnarray}
and
\begin{eqnarray}
  S^{(2)} &=&2\pi^2 \int d\xi\, \left[\rho^3 \left(\frac12 {\phi'}^2 + U \right)
    - \frac{3}{\kappa} \left( \rho{\rho'}^2+ \rho
    \right)\right] \cr
  &=& 4\pi^2 \int d\xi\, \left(\rho^3 U -  \frac{3}{\kappa} \rho \right) \, . 
\end{eqnarray}
In each case, the second equality holds for solutions of the field equations,
but not in general.   The difference between the two is
\begin{equation}
    S^{(2)}-S^{(1)} = -\frac{6\pi^2}{\kappa}\int d\xi \, \frac{d}{d\xi}
      \left(\rho^2 \rho'\right) \, .
\end{equation}  
This is in fact equal to the Gibbons-Hawking boundary term, so $S^{(2)}$ 
corresponds to the action as written in Eq.~(\ref{fullaction}).

For decay from a de Sitter vacuum, the bounce and the pure original
vacuum solutions are both compact and have no boundary term, so
$S^{(1)}-S^{(2)}$ clearly vanishes.  We can therefore restrict our
discussion to decay from a Minkowski or AdS false vacuum.  In both of
these cases the integrations extend out to $\xi = \infty$, and so may
be infinite.  To regulate these, we define $S^{(a)}(\bar \rho)$ to be
result of integrating only from $\xi=0$ to $\xi = \bar\xi$, where the
latter is defined by $\rho(\bar\xi) = \bar\rho$.  We then define the
tunneling exponent by
\begin{equation} 
    B^{(a)}= \lim_{\bar\rho \to \infty} B^{(a)}(\bar\rho) \equiv
    \lim_{\bar\rho \to \infty} \left[S^{(a)}_{\rm bounce}(\bar\rho)
    - S^{(a)}_{\rm fv}(\bar\rho) \right] \, .
\label{limits}
\end{equation}
We will show that this is finite for both values of $a$.  Furthermore,
we will show that that $B^{(1)}$ and $B^{(2)}$ are in fact equal or,
equivalently, that
\begin{equation}
   \lim_{\bar\rho \to \infty} \bar\rho^2 
    \left[ \rho'_{\rm  bounce}(\bar\xi) -\rho'_{\rm fv}(\bar\xi) \right] = 0  \, .
\label{rholimit}
\end{equation}

\subsection{Minkowski false vacuum}
    The pure false vacuum has $U=0$ and $\rho=\xi$.  It follows that 
$S^{(1)}_{\rm fv}(\bar\rho) =0$ and   
\begin{equation}
    S^{(2)}_{\rm fv}(\bar\rho) = - \frac{6\pi^2}{\kappa} \bar\rho^2 \, . 
\end{equation}
 
In the bounce solution at large $\xi$, $\phi$ approaches its false
vacuum value exponentially rapidly, so $U$ and $\phi'$ are both
exponentially small.  It then follows that $\rho'$ differs from unity
by an exponentially small amount.  Hence, if $\xi_1$ is well outside
the bounce, then for $\bar\xi > \xi_1$ we have
\begin{equation}
     \bar\rho=\rho(\bar\xi) = \rho(\xi_1) + \bar\xi -\xi_1 + \dots \, ,
\end{equation}
where the ellipsis represents exponentially small terms.  It is then 
straightforward to show both that $B^{(1)}(\infty)$ is
finite and that Eq.~(\ref{rholimit}) is satisfied.
 
\subsection{AdS false vacuum}
    For the AdS case, let us define $\Lambda= - U_{\rm fv}$ and
an AdS length 
\begin{equation}
      L = \sqrt{\frac{3}{\kappa \, |U_{\rm fv}|}} 
     = \sqrt{\frac{3}{\Lambda\kappa }} \, .
\end{equation}

The pure false vacuum solution is given by $\phi=\phi_{\rm fv}$ and
\begin{equation}
     \rho(\xi) = L \sinh(\xi/L) \, .
\end{equation}
It follows that 
\begin{eqnarray} 
    S^{(1)}_{\rm fv}(\bar\rho) &=& 2\pi^2 \Lambda \int_0^{\bar\xi} d\xi\, \rho^3 \cr
      &=& 2\pi^2 \Lambda L^4 \left[ \frac13 \cosh^3(\bar\xi/L) 
      -\cosh(\bar\xi/L) +\frac23 \right] \, .
\end{eqnarray}

At large distances the bounce solution can be written as 
\begin{eqnarray}
    \phi(\xi) &=& \phi_{\rm fv} + \delta\phi(\xi) \, ,  \cr
    \rho(\xi) &=& L \sinh(\xi/L + \Delta) + \delta\rho(\xi)  \, .
\end{eqnarray}
It is convenient to define $\eta = \xi/L + \Delta$. Then, to 
first order in $\delta\phi$, Eq.~(\ref{phieq}) becomes
\begin{equation}
    \frac{d^2 \delta\phi}{d\eta^2} + 3\coth \eta \,\frac{d\delta\phi}{d\eta}
           = C \delta\phi  \, ,
\end{equation} 
where 
\begin{equation}
    C = L^2 U''(\phi_{\rm fv})  \, .
\end{equation}

For $\xi \gg L$ we have 
\begin{equation}
     \frac{d^2 \delta\phi}{d\eta^2} + 3 \,\frac{d\delta\phi}{ d\eta}
     = C \delta\phi \, .
\end{equation}
The solution with $\delta\phi$ tending to zero as $\xi \to \infty$ 
is 
\begin{equation}
     \delta\phi  = Q \, e^{-a\eta} \, ,
\label{deltaPhiLim}
\end{equation}
where
\begin{equation}
     a = \frac32\left(1 + \sqrt{1+ \frac{4C}{9}}\right) \, .
\end{equation}

We can now investigate the convergence of the actions. With $\xi_1 \gg L$
well outside the bounce, and $\bar\rho > \rho_1$, we have 
\begin{eqnarray}
    B^{(1)}(\bar\rho) &=&  B^{(1)}(\rho_1) 
  -2\pi^2 \int_{\xi_1}^{\bar\xi} d\xi \,\rho^3 \left[U(\phi)-U_{\rm fv}\right] \cr
     &=& B^{(1)}(\rho_1)  -2\pi^2 \int_{\rho_1}^{\bar\rho} d\rho \,
             \frac{\rho^3}{\rho'} \,[U(\phi) - U_{\rm fv}] \cr
    &\approx& B^{(1)}(\rho_1)  -\pi^2 CL^{-1}\int_{\rho_1}^{\bar\rho} d\rho \,
               \rho^2 \delta\phi^2  \, .
\end{eqnarray}
For $\xi$ sufficiently large, $\rho \sim e^\eta$, so, using  
Eq.~(\ref{deltaPhiLim}), we see that $\delta\phi \sim \rho^{-a}$ and the integrand
behaves as
\begin{equation}
      \rho^{2-2a} = \rho^{-1-3\sqrt{1 + 4C/9} } \, .
\end{equation}
This falloff is fast enough to make the integral converge as $\bar\rho \to \infty$.

To verify that both actions lead to the same tunneling exponent, we first 
note that Eq.~(\ref{rhoeq}) gives
\begin{equation}
    \rho' = \left[ 1 + \frac\kappa3 \, \rho^2 \left(\frac12{\phi'}^2 -U\right)
            \right]^{1/2}  \, .
\end{equation}
For $\bar\rho\gg L$ and in the asymptotic region, 
\begin{eqnarray}
     \rho'_{\rm bounce} &\approx& \frac{\rho}{L} \left\{ 1 + \frac12 
   \frac{\left[{\delta\phi'}^2  - U''(\phi_{\rm fv}) \delta\phi^2  \right] 
     }{U_{\rm fv}} \right\}^{1/2}   \cr
      &\approx& \rho'_{\rm fv} + \frac{\rho}{4 L U_{\rm fv} } \,
       \left[{\delta\phi'}^2  - U''(\phi_{\rm fv}) \delta\phi^2  \right] \, .
\end{eqnarray}
It follows that the surface term from Eq.~(\ref{rholimit}) behaves as 
\begin{equation} 
    \bar\rho^2  \left[ \rho'_{\rm bounce}(\bar\xi) -\rho'_{\rm fv}(\bar\xi) \right]
     \sim {\bar \rho}^{3-2a}  \, .
\end{equation}
Because $a > \frac32$, this vanishes in the limit $\bar\rho \to \infty$, as 
required.

\section{Numerical methods}
\label{numerical-app}

In this appendix we describe the numerical techniques that we used for
the calculation of the bounce and its action.  The presence of very small
numbers causes some complications which we address here. 

Our task was to solve Eqs.~(\ref{phieq}) and (\ref{rhoeq}), subject to the
boundary conditions of Eq.~(\ref{bdy-cond}).  This can be done by a 
shooting code in Mathematica.  Given the initial values $\phi'(0)=0$
and $\rho(0)$, we choose a value for $\phi(0)$ and integrate
forward until one of the following conditions is met:
\begin{enumerate}
	\item $\phi$ passes the false vacuum.  This is an overshoot trajectory.
	\item $\phi'$ vanishes before $\phi_{\rm fv}$ has been reached.
          This is an undershoot trajectory.
	\item $\rho'$ vanishes.  If this occurs, $\rho$ will start to
          decrease and eventually reach a second zero, giving a
          compact solution.  This is unacceptable because 
          decay from Minkowski space requires a bounce 
          with infinite $R^4$ topology.  Because $\rho'$ can only
          vanish when $U(\phi)$ is positive (i.e., in the potential barrier
          before $\phi_{\rm fv}$), this should be viewed as a
          undershoot trajectory.
\end{enumerate} 
The desired bounce solution lies at the boundary between the overshoot
and undershoot regions, and can be found by bracketing this point with
successive trial values of $\phi(0)$

However, subtleties can arise.  In principle, the procedure described
above should work for finding all CDL solutions. However, it is not
uncommon for $\phi_0$ to be exponentially close to the true vacuum. If
the wall is at $\xi_{\rm wall}$, the value of $\phi_0-\phi_{\rm tv}$
is typically of order of $e^{-m \xi_{\rm wall}}$, where $m$ is a
characteristic mass scale of the problem, and $m \xi_{\rm wall}$ can
be large. Therefore, we cannot represent these numbers in a machine
with enough precision. However, this smallness paves the way for an
analytic approximation.

We saw in Sec.~\ref{analytic-sec} that linearizing about the true vacuum 
led to the approximate solution
\begin{equation}\label{approx3}
    \phi(\xi)= \frac{2 \phi_0 C_\alpha^{(3/2)}(\cosh \xi/\ell )}{(\alpha+2)(\alpha+1)}~.  
\end{equation}
Differentiating with respect to $\xi$ gives
\begin{equation}\label{approx4}
     \phi'( \xi)= \frac{3 \sinh ( \xi/\ell )
       C_{\alpha-1}^{(5/2)}[\cosh ( \xi/\ell)]}
     {\ell\,C_\alpha^{(5/2)}[\cosh (\xi/\ell )]}
    \,\, \phi(\xi)  \, .
\end{equation}
If the value of $\phi(0)$ is too small, we do not integrate from
$\xi=0$. Instead we choose a value $\bar\xi$ such that the
corresponding field $\phi(\bar\xi)$ from \eqref{approx3} is small
enough that the linearized approximation is still valid but not too
small for the machine precision. We then take $\phi'(\bar\xi)$ and
$\rho(\bar\xi)$ from equations Eq.~(\ref{approx4}) and
Eq.~(\ref{rhoAsSinh}) and numerically integrate the equations of
motion with these initial conditions.

\section{Decays from AdS vacua}
\label{Ads-app}

Our analysis can be easily extended to the case where both
the true and the false vacua are AdS spaces.  As a result,
the pseudo-energy is everywhere positive definite and, 
except in the extremely weak gravity limit, $\rho$ is an exponentially
growing function of $\xi$ throughout the transition region
connecting the vacua.  The second term dominates the right-hand side of 
Eq.~(\ref{RhoPrimeRho}) throughout Region~II.  We can then make the
approximation 
\begin{equation}
  \frac{\rho'}{\rho}= \sqrt{\frac{\kappa}{3} \left(\frac12 \phi'^2-U\right)} \, .
\end{equation}
This again yields Eq.~(\ref{ddphiEq}) for the scalar field, and our
shooting method for finding the critical value of gravity remains
valid.

\begin{table}
\caption{\label{adsads-data} Similar to Table~\ref{data1}, but for an
AdS to AdS decay with $U_{\rm fv}/v^4=-0.01$, $U_{\rm tv}/v^4=-0.1$, 
and $U_{\rm top}/v =-0.005$.
For this potential $\beta_{\rm cr}=1.7502$.}
\begin{ruledtabular}
\begin{tabular}{|c|ccccccccc|}
  $\beta$ &$S$&$\xi_{\rm wall}$& $\rho_{\rm wall}$& $\ell_{\rm AdS}$&$\Delta \xi_{\rm wall}$&$\Delta \rho_{\rm wall}$&  $\eta $ &  $\sigma_{\rm wall}$  & $b$ \\ \hline 
  0 & 129 & 6.47 & 6.47 & $\infty$ & 3.51 & 3.51 &0.0432 & 0.0159& 0.478  \cr 
  1.0000& 194& 6.93& 8.07& 5.48& 3.57& 4.22& 0.0320&  0.0163 & 0.462  \cr 
  1.5000 & 457& 8.09& 13.2& 3.65& 3.63& 6.61& 0.0133&  0.0167& 0.441 \cr
  1.7000 & 1511& 10.2& 28.9& 3.22& 3.67& 14.2& 2.19$\times 10^{-3}$&  0.0169 & 0.431 \cr
  1.7300 & 2758& 11.6& 45.5& 3.17& 3.67& 22.3& 6.84$\times 10^{-4}$&  0.0169& 0.429 \cr
  1.7450  & 6269& 13.7& 90.2& 3.14& 3.67& 44.1& 1.10$\times 10^{-4}$&  0.0169 &0.429 \cr
  1.7490 &1.46$\times 10^4$ & 16.0& 194& 3.13& 3.67& 94.6& 1.36$\times 10^{-5}$&  0.0169& 0.428 \cr
  1.7496 &2.20$\times 10^4$& 17.3& 286& 3.13& 3.67& 139& 4.65$\times 10^{-6}$  &  0.0169 & 0.428 \cr
  1.7498 & 2.86$\times 10^4$ & 18.1& 368& 3.13& 3.67& 179& 2.33$\times 10^{-6}$  &  0.0169&0.428  \cr 
  1.7500 &4.92$\times 10^4$& 19.7& 623& 3.13& 3.67& 304& 5.44$\times 10^{-7}$   & 0.0169 & 0.428 \cr
  1.7501 & 1.96$\times 10^5$& 24.0& 2439& 3.13& 3.67& 1191 & 1.26$\times 10^{-8}$  & 0.0169 & 0.428 \cr  
\end{tabular}
\end{ruledtabular}
\end{table}

Data for an AdS to AdS decay are shown in Table~\ref{adsads-data}. We
again find complete agreement between our simulation and the critical
value of $\beta$ obtained by shooting.

\begin{figure}
   \centering
   \includegraphics[width=3.2in]{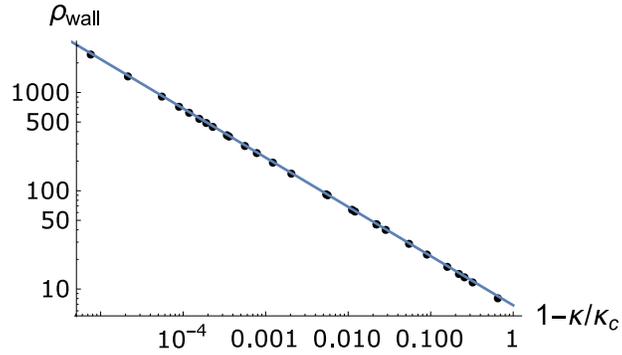}
   \caption{Behavior of $\rho_{\rm wall}$ (measured in
       units of $v^{-1}$) for the potential corresponding to the data
     in Table~\ref{adsads-data}.}
   \label{AdsAdsRho}
\end{figure}

\begin{figure}
   \centering
   \includegraphics[width=3.2in]{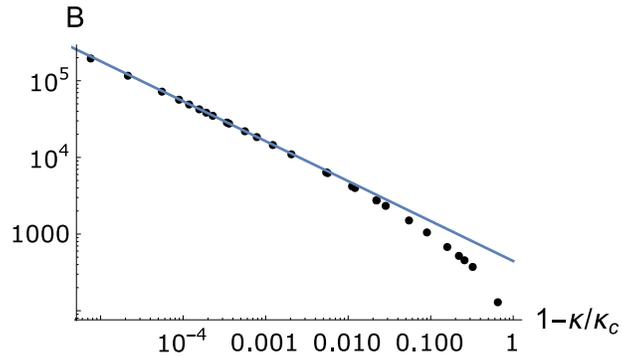}
   \caption{Behavior of $B$ for the potential corresponding to
the data in Table~\ref{adsads-data}.}
   \label{AdsAdsB}
\end{figure}

The behavior of the action and the wall radius as $\beta$ and $\kappa$ approach
their critical values is shown in Figs.~\ref{AdsAdsRho} and \ref{AdsAdsB}.
Near the critical values these are fit well by 
\begin{equation} \label{criticalBRho}
  \bar\rho  \sim (\kappa_{\rm cr} - \kappa)^{-1/2}~, 
          \qquad B  \sim (\kappa_{\rm cr} - \kappa)^{-1/2} \, .
\end{equation} 
This is in stark contrast with the decays from the Minkowski vacuum,
where $B \sim (\kappa_{\rm cr} - \kappa)^{-2}$ and $\bar\rho \sim
(\kappa_{\rm cr} - \kappa)^{-1}$. This can be understood by examining
Parke's thin-wall results for an arbitrary false vacuum~\cite{Parke:1982pm}.
If the false vacuum is Minkowski, the thin-wall analysis gives an
exact power law, with exponents $-2$ and $-1$, for all values of $\epsilon$.
If the false vacuum is AdS, there is approximate power law behavior,
with exponents $-1/2$ for both quantities, that is valid near the 
critical point.  The region where this approximation is valid shrinks
to zero as the false vacuum energy density approaches zero.

\end{document}